\documentclass[12pt,preprint]{aastex}

\newcommand{\nd}{\nodata}
\newcommand{\tblntmrk}{\tablenotemark}

\begin{document}

\title{Radio Sources Toward Galaxy Clusters at 30 GHz}
\author{
K.~Coble\altaffilmark{1,2,3}, 
M.~Bonamente\altaffilmark{4,5},
J.~E.~Carlstrom\altaffilmark{3,6}, 
K.~Dawson\altaffilmark{7},
N.~Hasler\altaffilmark{5},
W.~Holzapfel\altaffilmark{7},
M.~Joy\altaffilmark{4}, 
S.~LaRoque\altaffilmark{3},
D.~P.~Marrone\altaffilmark{8,3},
E.~D.~Reese\altaffilmark{9}
}

\altaffiltext{1}{Dept.\ of Chemistry and Physics, Chicago State University, 9501 South King Drive, Chicago, IL 60628, USA, coble@oddjob.uchicago.edu}
\altaffiltext{2}{NSF Astronomy and Astrophysics Postdoctoral Fellow, Adler Planetarium and Astronomy Museum, Chicago, IL, 60605, USA}
\altaffiltext{3}{Kavli Institute for Cosmological Physics, Dept.\ of Astronomy and Astrophysics, University of Chicago, 5640 South Ellis Ave., Chicago, IL 60637, USA}
\altaffiltext{4}{Dept.\ of Space Science - NSSTC, NASA Marshall Space Flight Center, Huntsville, AL 35812, USA}
\altaffiltext{5}{Dept.\ of Physics, University of Alabama in Huntsville, Huntsville, AL 35805, USA}
\altaffiltext{6}{Dept.\ of Physics, Enrico Fermi Inst., University of Chicago, 5640 South Ellis Ave., Chicago, IL 60637, USA}
\altaffiltext{7}{Dept.\ of Physics, University of California, Berkeley, CA 94720, USA, (KD now at LBNL)}
\altaffiltext{8}{Jansky Fellow, National Radio Astronomy Observatory}
\altaffiltext{9}{Physics Department, University of California, Davis, CA 95616, USA}


\begin{abstract}

Extra-galactic radio sources are a significant contaminant in cosmic
microwave background and Sunyaev-Zel'dovich effect experiments. 
Deep interferometric observations with the BIMA and
OVRO arrays are used to characterize the spatial, spectral, and flux
distributions of radio sources toward massive galaxy clusters at 28.5
GHz.  We compute counts of mJy source fluxes from 89 fields
centered on known massive galaxy clusters and 8 non-cluster
fields.
We find that 
source counts in the inner regions of the cluster fields (within 0.5
arcmin of the cluster center) are a factor of $8.9^{+4.3}_{-2.8}$
times higher than counts in the outer regions of the cluster fields
(radius greater than 0.5 arcmin). Counts in the outer regions of the
cluster fields are in turn a factor of $3.3^{+4.1}_{-1.8}$ greater
than those in the non-cluster fields. Counts in the non-cluster fields
are consistent with extrapolations from the results of other
surveys.
We compute spectral indices of mJy sources in cluster fields
between 1.4 and 28.5 GHz and find a mean spectral index of $\alpha
= 0.66$ with an rms dispersion of $0.36$, where flux $S \propto \nu^{-\alpha}$.
The distribution is skewed, with a median spectral index of $0.72$
and 25th and 75th percentiles of $0.51$ and $0.92$, respectively.
This is steeper than the spectral indices of stronger 
field sources measured by other surveys.

\end{abstract}



\section{INTRODUCTION}

Extra-galactic radio sources are a significant contaminant in cosmic
microwave background (CMB) and Sunyaev-Zel'dovich effect (SZE)
experiments (e.g., see \citet{holder2002, knox2004, tegmark2000}). Measurements of
the CMB and SZE
\citep{sunyaev1970,sunyaev1972} have the potential to yield a wealth 
of cosmological information if foreground contaminants are well-understood.
Models for the number counts of radio sources
as a function of flux \citep{dezotti2005, toffolatti1999, sokasian2001} have been
derived from observations at lower frequencies and extrapolated to
microwave frequencies.

Radio sources are often associated with the clusters of galaxies
themselves. This is a potential source of bias for current and planned
SZE surveys such as the
Sunyaev-Zel'dovich Array (SZA)\footnote{SZA website:
http://astro.uchicago.edu/\~sza}, the Arcminute Microkelvin Imager
(AMI)\footnote{AMI website: http://www.mrao.cam.ac.uk/telescopes/ami/}, the
Atacama Pathfinder Experiment Sunyaev-Zel'dovich (APEX-SZ)
survey\footnote{APEX-SZ website: http://bolo.berkeley.edu/apexsz}, the South
Pole Telescope (SPT)\footnote{SPT website: http://spt.uchicago.edu/},
and the Atacama Cosmology Telescope (ACT)\footnote{ACT website:
http://www.hep.upenn.edu/act/}.
To understand the impact on planned SZE cluster surveys, it is critical to 
characterize the spatial, spectral, and flux distribution of 
sources associated with clusters.
 
Contaminating radio emission from extra-galactic sources 
at frequencies  less than approximately 100 GHz is attributed to synchrotron radiation
from active galactic nuclei (AGN) and star-forming galaxies.
The AGN-powered
radio galaxies dominate the source counts at high luminosities.
At higher frequencies, the emission is attributed to
dust emission from star-forming galaxies.
Observations of radio sources toward galaxy clusters
at very low frequencies ($\lesssim 5$ GHz; e.g., \citet{slee1983,slee1998,
owen1996, ledlow1995, reddy2004}) show a strong
central concentration of radio galaxies in clusters. The distribution of 
synchrotron-emitting star-forming galaxies
is found to be less centrally-peaked in clusters \citep{rizza2003}
than that of AGN-powered radio galaxies.
At our sensitivity level and observing frequency of 28.5 GHz, radio
sources powered by AGN dominate and we refer to them henceforth 
as radio sources.

Several CMB experiments, such as WMAP \citep{bennett2003}, DASI
\citep{kovac2002}, VSA \citep{cleary2005}, and CBI \citep{mason2003},
have measured microwave source counts as a function of flux for sources
brighter than about 10 mJy.
There have been two prior analyses of
radio sources in SZE data taken with the Owens Valley Radio Observatory 
(OVRO)\footnote{The OVRO mm-wave array is operated by Caltech with support
from the National Science Foundation}
and Berkeley-Illinois-Maryland Association
(BIMA)\footnote{The BIMA array is operated with support
from the National Science Foundation} arrays.
Using a sample of 56 fields centered on known massive galaxy clusters,
\citet{cooray1998a} computed counts and spectral indices of
radio sources.  Using the outer regions of 41 cluster fields,
\citet{laroque2002} computed the normalization of source counts as a
function of flux for faint sources in SZE data. The data used in
\citet{cooray1998a} and in \citet{laroque2002} are subsets of the data
presented in this paper.

In this paper we analyse faint ($\sim$~mJy) radio sources found
serendipitously toward massive galaxy clusters at 28.5 GHz from the
OVRO/BIMA SZE imaging project.  Characterizing the spatial, spectral,
and flux distribution at this relatively high frequency should help
improve projections for radio source contamination in SZE and CMB
experiments at frequencies of 30 GHz and higher.  We use 89 fields
centered on known massive galaxy clusters and 8 non-cluster
fields. The paper is organized as follows: Section 2 reviews the
observations, data reduction, field selection, and the measured source
fluxes. In Section 3, we compute spectral indices between 1.4 and 28.5
GHz using fluxes from our data and published 1.4 GHz surveys. In
Section 4, we present source counts as a function of flux for cluster
and non-cluster fields as well as the angular radial dependence
of counts in cluster fields. We compare our results with those from
other experiments and with theoretical models.


\section{OBSERVATIONS}

\subsection{Observations}

The 28.5 GHz observations were carried out with BIMA during the summers of
1996 -- 2002 and OVRO during the
summers of 1995 -- 2001 as part of the OVRO/BIMA SZE imaging project
(see, for example, \citet{reese2002} and \citet{grego2001}). A total
of 62 cluster fields were observed at BIMA and 55 cluster fields were
observed at OVRO. A total of 28 cluster fields were observed at
both BIMA and OVRO, yielding observations of 89 unique cluster
fields. For the BIMA CMB finescale anisotropy project \citep{dawson2002,dawson2006},
a total of 18 non-cluster fields were observed.

The BIMA array consists of ten 6.1 meter diameter telescopes
with primary beams of 6.6 arcmin FWHM; nine of the ten BIMA
telescopes were used for the 28.5 GHz observations. The OVRO array consists 
of six 10.4-meter telescopes, with primary beams of 4.2 arcmin FWHM; all six
OVRO telescopes were used. 
The primary beams were measured holographically and were found to
be well-approximated by Gaussians; we use the azimuthally averaged
measured beam responses in our analyses. For CMB observations at BIMA,
the array was set in a compact configuration to maximize brightness
sensitivity. For cluster observations at both OVRO and BIMA, most of the
telescopes were also configured in a compact configuration that provided dense $u-v$ coverage
to the shadowing limit, and one or two telescopes were
placed at longer baselines for higher angular resolution monitoring of sources.
The longest baselines used at OVRO and BIMA ranged from 70 to 140 meters
for the data presented here.

The telescopes were outfitted with cm-wave receivers
\citep{carlstrom1996} equipped with cryogenically cooled 26 - 36 GHz
HEMT amplifiers \citep{pospieszalski1995}.  Typical receiver
temperatures were 11~K to 20~K, and 
when integrated
with the OVRO and BIMA telescopes yielded typical system temperatures 
scaled to above the atmosphere of 45~K to
55~K, and as low as 35 K.
OVRO observations were correlated with an analog correlator consisting of two 1 GHz bandwidth channels centered at 28.5 and
30 GHz. The $u-v$ data from the two channels were not combined before making maps or fitting sources to the data. The OVRO correlator integration time was four minutes or less. BIMA observations were correlated with a multi-channel hybrid correlator. After calibration, the $u-v$ data were reduced to a single 0.8 GHz wide bandwidth centered at 28.5 GHz. The BIMA correlator integration time was 50 seconds.

Observations of cluster and CMB fields were interleaved every $\sim$ 20
minutes with observations of a strong point source for phase
calibration. Source data that were not bracketed in time by phase calibrator data were discarded, as were data from baselines in which one of the array elements was
shadowed by another.  Source data were also discarded if the bracketing phase
calibrator observations showed a discontinuity in the instrument phase response. Lastly, source data were discarded if atmospheric phase fluctuations showed a loss of correlation greater than a few percent on the long baseline observations of the phase calibrator.
Observations of the phase calibrators indicate the gain
stability was stable to $\sim 1 \% $ over several months. 
The absolute
calibration is based on observations of Mars, with the brightness
temperature taken from the \citet{rudy1987} model.  Further details
of the observations and data reduction can be found in 
\citet{grego2001} and \citet{reese2002}.

\subsection{Field Selection}

The cluster fields of the OVRO/BIMA SZE imaging project were chosen 
to obtain precise measurements of the SZE in massive galaxy clusters. 
Potential targets were screened for strong radio
sources using archival data at lower
frequencies such as NVSS \citep{condon1998} and FIRST \citep{white1997}.
In addition, if a strong source ($>10 - 20$ mJy) was
detected near the cluster center
in the initial 28.5 GHz observations, observations ceased in
favor of other less contaminated targets.
Due to the constraints of this field selection,
we do not attempt to characterize the distribution of
sources brighter than 10 mJy in this analysis.

The cluster fields were chosen mainly from X-ray catalogs and one optical
survey, including the following: (1) the ROSAT Brightest Cluster Survey, BCS
\citep{ebeling1997,ebeling1998,ebeling2000a,crawford1999}, (2) the
Einstein Observatory Extended Medium Sensitivity Survey, EMSS
\citep{gioia1990, stocke1991, gioia1994, maccacaro1994}, (3) the ROSAT
X-ray Brightest Abell Clusters, XBACS
\citep{ebeling1996,ebeling1996b}, (4) the Wide Angle ROSAT Pointed
Survey, WARPS \citep{scharf1997,jones1998,fairley2000, ebeling2000b},
(5) the Massive Cluster Survey, MACS \citep{ebeling2001}, and (6) the
Red-Sequence Cluster Survey, RCS \citep{gladders2005}.

Clusters were selected from the BCS, EMSS, and XBACS X-ray catalogs on
the basis of high X-ray luminosity.  Clusters from WARPS and MACS
were selected for X-ray luminosity and high redshift.  The 8
clusters selected for SZE observations from MACS form a
complete redshift-selected flux-limited X-ray sample and were chosen
regardless of possible radio source contamination \citep{laroque2003}.
A few optically selected clusters were also provided by the RCS
team. Table 1 lists the coordinates of the pointing centers for the
cluster fields.

The BIMA array was also used to observe 18 non-cluster fields for CMB
anisotropy measurements \citep{dawson2002,dawson2006}. The pointing
center coordinates for these fields are provided in Table 2. Of these
fields, 
only the 8 fields BDF14-BDF21 \citep{dawson2006} were chosen without regard to
possible radio source contamination. 
The fields BDF4, and BDF6-BDF13 \citep{dawson2002}
were chosen from NVSS to have minimal
contamination from strong radio sources.
The field HDF is centered on the Hubble Deep Field and was also selected 
to have no strong radio sources.
Only the 8 fields selected without regard to source contamination
are used in the analysis presented in this paper.

\subsection{Source Fluxes}

The positions and fluxes of 28.5~GHz sources and SZE decrements are
determined using the DIFMAP software package \citep{pearson1994}. The
SZE data consist of positions in the Fourier domain (also called the
$u-v$ plane) and the visibilities---the complex Fourier component
pairs as functions of $u$ and $v$, which are the Fourier conjugate
variables to right ascension and declination. DIFMAP is used to create
a map from the $u-v$ data using all baselines and natural weighting
($\propto \sigma^{-2}$). Source and decrement positions are
determined from this map. Emission sources are initally assumed to be
pointlike, while the SZE decrement (if any) is approximated with an
elliptical, isothermal $\beta$ model. The fourier transforms of the
model components are jointly fit to the observed visibilities. The
image of the residuals is searched for additional sources at greater
than five times the image rms and any sources found in this way are
added to the model and jointly refitted. Noise estimates are extracted
from the final residual images.

The source fluxes determined in this way depend on the coverage of the
$u-v$ plane for sources with structure on scales larger than that
probed by the longest baselines (30 arcsec to 15 arcsec for the 70 to
140 meter maximum baselines employed here). To enable a uniform
comparison with lower frequency data from the NVSS, which has an
angular resolution of 45 arcsec, we repeat the $u-v$ fitting procedure
considering only baselines shorter than 4~$k\lambda$. This provides
matched angular resolution at a cost of typically $40-50$\% of our
data, yielding a sensitivity loss of $30-40$\%. We also measured the
flux within an extended area centered on each radio source in the CLEANed 
maps made from the 4~$k\lambda$ cut and full $u-v$ data set.
We compare the fluxes
obtained from the $u-v$ model fits and the analysis of the CLEANed maps 
from both 
the cut and the full $u-v$ data. We find the fluxes found with the four methods
agree for 103 out of 121 sources, double-counting sources detected with
both arrays because of the difference in sensitivity and $u-v$
coverage. In these cases we conclude that the point source
approximation is adequate for our angular resolution and sensitivity
and we report the fluxes and noise levels determined from all
baselines in Tables 1 and 2. The remaining 18 sources have best-fit
point source fluxes that depend on our choice of $u-v$ range. We
find, however,  that the total flux recovered by the CLEAN algorithm in the
region around each source in the full data set matches that obtained
for the flux found with only the short baselines for all but one of
these sources. In these cases we use the
flux measured in the CLEANed map made from the full data set. Finally,
in one case (MACS~J0717.5+3745, source 1) we do not recover the full
short-baseline flux when CLEANing the image of the full data set. In
this case we have used the flux and noise measured for just the
baselines shorter than 4~$k\lambda$ for both sources in the field.

Table 1 lists the 28.5 GHz beam attenuation-corrected fluxes for
sources detected in the BIMA and OVRO cluster fields and the centroid
of the SZE decrement, if any. Noise is reported for the field center in
column 7, while the uncertainty in the flux of each source (column 15)
properly accounts for the beam profile. The radial distance from the
pointing center, or the SZ decrement if observed, to each source is
also provided. At the 5$\sigma$-level or greater we detect 62 sources in the 62 BIMA
cluster fields and 55 sources in the 55 OVRO cluster fields.  A
total of 22 sources are detected at both BIMA and OVRO, yielding 95
unique sources in the 89 cluster fields. For sources observed at both
BIMA and OVRO, the flux measurements are in good agreement.  Source
fluxes, positions, and noise levels for the non-cluster fields are
given in Table 2. We detect two sources at $\ge 5 \sigma$
in the 8 BIMA non-cluster fields which were selected without
regard to possible source contamination.

The fluxes in Tables 1 and 2 do not account for attenuation due to
temporal and spectral averaging of the $u-v$ data, effects which are
far less significant than the beam attenuation. Averaging of
interferometric data in the $u-v$ plane leads to attenuation of the
amplitude response as a function of angular distance from the field
center that depends on the telescope array configuration. However,
these effects have been restricted to insignificant levels as part of
the experimental design. The bandwidth constraint translates to
${\Delta \nu \over \nu} \times {B \over D} < 1 $, where ${\Delta \nu
\over \nu}$ is the fractional bandwidth, $D$ is the diameter of an
array element and $B$ is the baseline. For the longest baselines we
find values of ${\Delta \nu \over \nu} \times {B \over D}$ to be 0.32
- 0.64 and 0.24 - 0.47 for the BIMA and OVRO observations,
respectively. For measurements using only short baselines
($<4k\lambda$) these values decrease to 0.18 and 0.13, respectively.
The correspondence between our measured fluxes using all baselines and
just the short baselines also suggests that bandwidth smearing is not
important. The integration time constraint translates to ${2\pi
\over 24} t_{int} \times {B \over D} < 1 $, where $t_{int}$ is the
$u-v$ averaging time in hours. For the longest baselines and
integration times we find values of ${2\pi \over 24} t_{int} \times {B
\over D}$ to be less than 0.0035 and 0.001 for the BIMA and OVRO
observations, respectively.


\section{SPECTRAL INDICES}

We use the results of surveys at lower observing frequency to constrain
the spectral indices of radio sources detected with the BIMA/OVRO observations.
Fluxes at 1.4 GHz are taken primarily from the NVSS catalog, which has a
resolution of 45 arcsec and limiting peak source brightness of 2.5 mJy.
We obtain 1.4 GHz fluxes from the FIRST catalog (limiting flux of 1 mJy
and resolution of 5 arcsec) for several additional sources which 
were below the NVSS detection threshold.
We obtain 1.4 GHz fluxes from VLA archival
maps for several sources which were not in the NVSS or FIRST catalogs.
Of the 95 sources in the cluster fields, 88 have unambiguous
counterparts in the NVSS, FIRST, or VLA archival data.
All 28.5 GHz sources within 0.5 arcmin of the cluster centers were confirmed
to have counterparts at 1.4 GHz.
The 1.4 GHz fluxes and the results of a literature search 
for low frequency measurements of
sources not identified in the NVSS, FIRST, or VLA surveys
are given in Table 1. Unless otherwise noted, the tabulated 
1.4 GHz fluxes are from NVSS.  
For the 7 sources that do not have unambiguous 1.4 GHz counterparts
in NVSS, FIRST, or the VLA archive, we assume a 1.4 GHz flux equal to 
three times the survey noise.
In reality, the true 1.4 GHz fluxes typically will be weaker and the true spectral 
indices of these sources more shallow.
However, because there are only 7 sources lacking 1.4 GHz detections,
the bias in the resulting mean spectral index for all sources
in cluster fields computed below is small.
Furthermore, because the 1.4 and 28.5 GHz measurements were
not made contemporaneously, variability of the sources
may contribute to a broadening in the distribution of spectral indices.

We compute spectral indices between 1.4 and 28.5 GHz for sources
in the cluster fields (selected at 28.5 GHz) where index
$\alpha$ is defined by $S \propto \nu^{-\alpha}$.
If we use only the 88 sources with detections in NVSS, FIRST, or the VLA maps
and omit the 7 sources lacking 1.4 GHz detections,
the resulting mean spectral index is $\alpha = 0.66$ with an rms dispersion of $0.36$.
If we include limits for the 7 sources lacking 1.4 GHz detections,
the mean spectral index $\alpha = 0.60$ with an rms dispersion of $0.42$,
and a median index of $0.71$,
indicating that the bias due to omitting these 7 sources
is small. We use the 
88 sources with NVSS, FIRST, or VLA counterparts for the
remainder of our spectral index analysis.
A histogram of spectral indices for these 88 sources is shown in Figure 1.
Characterizations of the spectral index distribution are given in Table 3.
The distribution has a tail at low-$\alpha$
and is not well fit by a Gaussian. We therefore also compute
the median of the distribution, as well as the 25th and 75th
percentiles and find them to be $0.72$, $0.51$, and $0.92$,
respectively. 

While the beam attenuation
factor is potentially a significant source of uncertainty for
the 28.5 GHz sources, we find that excluding sources at large radii
(attenuation factor greater than 5.0) does not change the results.
We choose a maximum cutoff outer radius of 6.6 arcmin for BIMA and 4.2 arcmin for OVRO,
corresponding to a beam attenuation factor of about 30 and
spanning a region twice the FWHM of the primary beam.
When a source has observations from both BIMA and OVRO, we
choose the one with the best combination of sensitivity and survey
area.

The radial distribution of spectral indices is shown in Figure 2;
there is no apparent trend in spectral index with radius from the
cluster center.
We compare the spectral index distribution of the central regions of
cluster fields ($r < 0.5$ arcmin) with the distribution of the 
outer regions of cluster fields ($r > 0.5$ arcmin) and find no significant
differences.
The mean spectral index for the central regions of
cluster fields is $\alpha = 0.75$ with an rms dispersion of $0.24$ and the
mean spectral index for the outer regions of cluster fields
is $\alpha = 0.63$ with an rms dispersion of $0.38$. The medians
are $0.76$ [0.56 (25\%), 0.94 (75\%)] for the central regions
and $0.71$ [0.42 (25\%), 0.88 (75\%)] for the outer regions.
The results of a Kolmogorov-Smirnov (KS) test indicate that
the distribution of spectral indices of sources in the
inner regions of the cluster fields is consistent with
that of the outer regions. The maximum distance between
their cumulative distribution functions is 0.18, corresponding to
an $64\%$ probability that the two samples are drawn from the same distribution.
Since we have few sources in non-cluster fields, we do not compute
an average spectral index for this group. 

We compare the spectral index distribution of our mJy
cluster sources to those of somewhat brighter field sources measured by
other groups \citep{mason2003,waldram2003, bolton2004};
a summary is given in Table 3.
The CBI group \citep{mason2003} finds a spectral index from 1.4 to 31 GHz
of $0.45$ with an rms dispersion of $0.37$.
As a follow-up to the 9C survey \citep{waldram2003}, \citet{bolton2004} compute
indices between several frequencies from 1.4 to 43 GHz, distinguishing
between weak and strong sources.
The mean spectral index for the lower flux sample is $\sim 0.4$ for
the lower frequencies, and steepens to $\sim 0.9$ from 15.2 to 43 GHz.
\citet{waldram2003} and \citet{bolton2004} also find that a
greater percentage of the strong sources have a flat or rising spectrum.

Our spectral indices are somewhat steeper than the spectral indices
measured by CBI and much steeper than those of the $\sim$~Jy 
sources ($\alpha \sim 0$) measured by WMAP \citep{bennett2003}.
When comparing source surveys, it is essential to consider 
the flux and frequency at which the sources are selected.
We expect a survey of strong sources selected at high frequency 
to have a flatter spectral index than a survey of low flux sources 
selected at lower frequency. 
Our results are for a relatively low flux survey selected at high 
frequency, and it is interesting that the spectral index is relatively 
steep.
The sources in our survey primarily lie in the environments of rich galaxy
clusters and it is possible that there are significant differences 
between this population of sources and that found in surveys
that do not target clusters.
Our spectral indices for radio sources towards clusters
are similar to those found towards clusters at 2.7 GHz by \citet{slee1983}.
They find that spectral indices are steeper in clusters than in the field
and note a trend of shallower spectral indices with increasing cluster radius. 
Using our mean spectral index of 0.66, the lower flux limit of 0.12 Jy at 2.7 GHz
from \citet{slee1983} translates to 25 mJy at 28.5 GHz, which is slightly stronger
than the upper limit of our source sample.

As we look at sources with higher redshift, the emission frequency of
the radiation increases.  We might, therefore, expect these sources to
have steeper spectral indices.  However, the sources are selected
at higher frequency which might bias the sample toward flatter
spectral indices.  In Figure 3, we plot spectral index as a function
of cluster redshift and see no clear trend.  The mean and rms
dispersion in the spectral index are 0.67 and 0.37 for $z < 0.5$, compared to
0.64 and 0.29 for $z > 0.5$, and 0.76 and 0.20 for $z > 0.8$.


\section{SOURCE COUNTS}

\subsection{Analysis}
\label{sec:analysis}

With the field selection effects in mind from Section 2.2, we compute the
differential source counts as a function of flux,
$dN/dS$, in several flux bins, accounting for the varying noise levels from field to field.
We chose the flux bins in order to maximize the number of sources used and to
have a similar number of sources in each bin.

The survey boundary of each field for a given flux bin is set by the noise level of the field.
For each flux bin and field, the minimum level in the flux bin sets the allowable beam-corrected
noise level and the corresponding maximum attenuation radius for the field.  
For example, for $\ge 5 \sigma$ sources in a flux bin of 1.5 - 2.5 mJy,
the allowable beam-corrected noise level is $1.5/5 = 0.3$ mJy. 
This noise level sets the attenuation radius for the field, the radius at which the beam attenuation factor 
equals the beam-corrected noise level divided by the uncorrected noise level.
We set an outer boundary on the survey area for the field using the lesser of the attenuation radius or a maximum
cutoff outer radius away from the field pointing center.
We choose a maximum cutoff outer radius of 6.6 arcmin for BIMA and 4.2 arcmin for OVRO,
corresponding to a beam attenuation factor of about 30 and spanning a region twice the FWHM of the primary beam.
We treat this as a hard maximum cutoff; even if the noise is sufficiently low to allow us to go to
greater radii in our sampling of a field, we do not.
The outer boundary is measured relative to the field pointing center.

We further break the data into radial bins from the cluster center. The
cluster center is determined by the location of the SZE decrement. For
fields without a SZE decrement detection, the pointing center is used
as the center of the field. For each field, flux bin,
and radial bin, we compute the survey area within the boundary set by
the radial bin and the noise level for the field. Typically the survey
region for a given field, flux bin, and radial bin is a circle or
annulus, sometimes cut off by the noise boundary. We compute the total
survey area for each flux bin by adding up the area in all the
fields. When a field has observations from both BIMA and OVRO, we
choose the one with the best combination of sensitivity and survey
area.

For each field we identify all $\ge 5 \sigma$ sources in the survey area
that fall between the minimum and maximum fluxes of each flux bin.
We count up the sources in each flux bin to get raw total
source counts in the total survey area.
The errors for the raw counts in each bin are assumed to be Poisson distributed.
Differential source counts ($dN/dS$) and the associated errors 
are calculated by dividing the total raw counts in each bin
by the total survey area for the corresponding flux bin 
and by the flux bin width.

\subsection{Results and Discussion}

Differential source counts ($dN/dS$), the number of sources,
and the survey area for each flux bin are given in Figure 4 and Table 4 for
the central regions of the cluster fields (r $\leq 0.5$ arcmin),
the outer regions of the cluster fields (r $\geq 0.5$ arcmin),
and for the non-cluster fields.
The error bars on $dN/dS$ are the Poisson errors on the raw source counts and
do not include other sources of uncertainty.
Typical raw counts of sources are $\sim 4$ in each flux bin for the inner
regions of the cluster fields and $\sim 8$ for the outer cluster regions.
We only detect two $\ge 5 \sigma$ sources in the 8 non-cluster fields
that were selected without regard to possible radio source contamination.

The differential source counts can be described by a power law,
$d{\rm N}({\rm S})/d{\rm S} = N_0 (S/S_0)^{-\gamma}$, where $S_0 = 1\,$mJy
for this analysis.
Best fits using a Markov chain algorithm that
simultaneously estimate the normalizations for the inner, outer, and non-cluster regions, 
and a common power-law index
are shown with the data in Figure 4 and are given in Table 5.
The best-fit common power-law index is $\gamma=-1.98 \pm 0.20$.
Best fit power-laws for the central and outer cluster regions individually
are also shown in Figure 4 and Table 5.
As a cross-check, we compute $dN/dS$ for the BIMA and OVRO fields
separately and find good agreement; see Figure 5.
All uncertainties represent 68\% confidence intervals unless otherwise noted. 

Source counts are found to be greatly elevated toward the
central core of the cluster fields.
Using the normalizations from the best simultaneous fit,
source counts are found to be a factor 
of $8.9^{+4.3}_{-2.8}$ higher in the central
regions than in the outer regions of the cluster fields.
Counts are also elevated in the outer regions of the cluster fields relative
to the non-cluster fields by a factor of $3.3^{+4.1}_{-1.8}$.
These overdensities imply that $97^{+2}_{-5} \%$ of sources in the inner regions
are cluster members, as are $70^{+16}_{-37} \%$ of sources in the outer regions.
A comparable overabundance of radio sources toward galaxy clusters
is also seen at lower radio frequencies (e.g., \citet{slee1983,slee1998, owen1996, ledlow1995, reddy2004, rizza2003}).

We considered the possibility 
that gravitational lensing of background radio galaxies 
could produce the overabundance of detected radio
sources in the direction of massive galaxy clusters.  
A gravitational lens with magnification factor $\mu$ will modify the source counts
to $d{\rm N}^{\prime}({\rm S})/d{\rm S} = 
(d{\rm N}({\rm S/\mu})/d{\rm S}) / \mu^2$.
If the unlensed source counts can be described by a power law, 
$d{\rm N}({\rm S})/d{\rm S} \propto S^{-\gamma}$, then the source counts
will be changed by a factor $B= (d{\rm N^{\prime}}({\rm S})/d{\rm S}) / 
(d{\rm N}({\rm S})/d{\rm S}) = \mu^{-2+\gamma}$ 
\citep{blain2002}. 
The mean magnification in the BIMA and OVRO cluster fields is estimated by 
\cite{cooray1998a} to be $\mu \sim 1.4$. 
In this analysis, we temporarily assume that all sources are background sources
drawn from the same distribution and capable of being lensed.
Using the best joint fit power law index of $\gamma=1.98 \pm 0.20$ 
we expect a factor of $B = 0.99^{+0.07}_{-0.06}$, i.e.,
no change in source counts in the direction of the clusters due to lensing.
Therefore we conclude that, regardless of the magnification, gravitational 
lensing can not be responsible for the significant excess of sources seen in 
the direction of clusters. 

In Figure 6 we compare our measurements of $dN/dS$ with the \citet{dezotti2005}
30 GHz model and with measurements from other experiments, including WMAP, DASI, VSA, and CBI,
which all examine non-cluster fields.
We present source counts in terms of $log_{10}(S^{5/2} dN/dS)$, for ease of comparison.
Counts in our non-cluster fields are consistent with those expected from the model
and from extrapolations from other experiments, although with only two $\ge 5 \sigma$
sources in those fields, the sample variance is large.
The source counts toward cluster fields have a similar power-law slope,
but a higher normalization than expected from extrapolations of measurements 
of sky not concentrated on clusters.


\section{CONCLUSIONS}

From deep interferometric observations at 28.5 GHz of unresolved radio sources toward 89 fields centered on
massive galaxy clusters and 8 non-cluster fields,
we find that differential source counts are greatly elevated in the centers of cluster fields.
Counts are a factor of $8.9^{+4.3}_{-2.8}$ higher in central regions
(r $\leq 0.5$ arcmin) than in the outer regions (r $\geq 0.5$ arcmin) of the cluster fields.
Counts in our non-cluster fields are consistent with those expected from models and
from extrapolations from other experiments.
In addition, source counts in the outer regions of cluster fields are a factor of
$3.3^{+4.1}_{-1.8}$ higher than counts in non-cluster fields.

Using the NVSS and other surveys, we find a mean spectral index 
for sources in cluster fields between 1.4 and $28.5\,$GHz of $\alpha = 0.66$ 
with rms dispersion of $0.36$, where flux $S \propto \nu^{-\alpha}$. 
The distribution is skewed, with a median spectral index of $0.72$
[$0.51$ (25\%), $0.92$ (75\%)].
This is steeper than spectral indices of stronger ($\sim$~20~mJy)
field sources, and much steeper than
those of much stronger ($\sim$~Jy) field sources measured by other surveys.
No significant differences are found between the distributions of the spectral indices of sources in the inner and outer regions of the clusters.

These results can be used for improving forecasts for radio source contamination of
SZE and CMB experiments.
The cluster fields used in this work were chosen to contain massive clusters
and the sources were identified at $28.5\,$GHz.
We anticipate that the SZA and other instruments will be able to extend
this work to less massive clusters and to sources identified at yet higher
frequencies.


\acknowledgments

We gratefully acknowledge the excellent support of the
BIMA and OVRO staff over the many years of the OVRO/BIMA SZE program,
including J.R.\ Forster, C.\ Giovanine, R.\ Lawrence, S.\ Padin, R.\
Plambeck, S.\ Scott and D.\ Woody.  We thank C.\ Alexander,
A.\ Cooray, L.\ Grego, G.\ Holder, A.\
Miller, J.\ Mohr, S.\ Patel and P.\ Whitehouse for their contributions
to the SZE instrumentation, observations, and analysis.  
We thank G. de Zotti for providing the model of differential source counts.
We thank H. Andernach and the anonymous referee for several insightful suggestions which improved the paper.

This work was supported in part by NASA LTSA grant NAG5-7985, 
NSF grants PHY-0114422 and AST-0096913, the David and Lucile
Packard Foundation, the McDonnell Foundation, and a MSFC director's
discretionary award.  Research at the Owens Valley Radio Observatory
and the Berkeley-Illinois-Maryland Array was supported by NSF
grants AST 99-81546 and 02-28963. 
KC was supported by NSF grant AST-0104465 under the Astronomy
and Astrophysics Postdoctoral Fellowship program while at the University of Chicago
and the Adler Planetarium and Astronomy Museum.
SL
acknowledges support from the NASA Graduate Student Researchers
Program.


\bibliographystyle{apj}



\clearpage

\begin{figure}
\epsscale{1.0}
\plotone{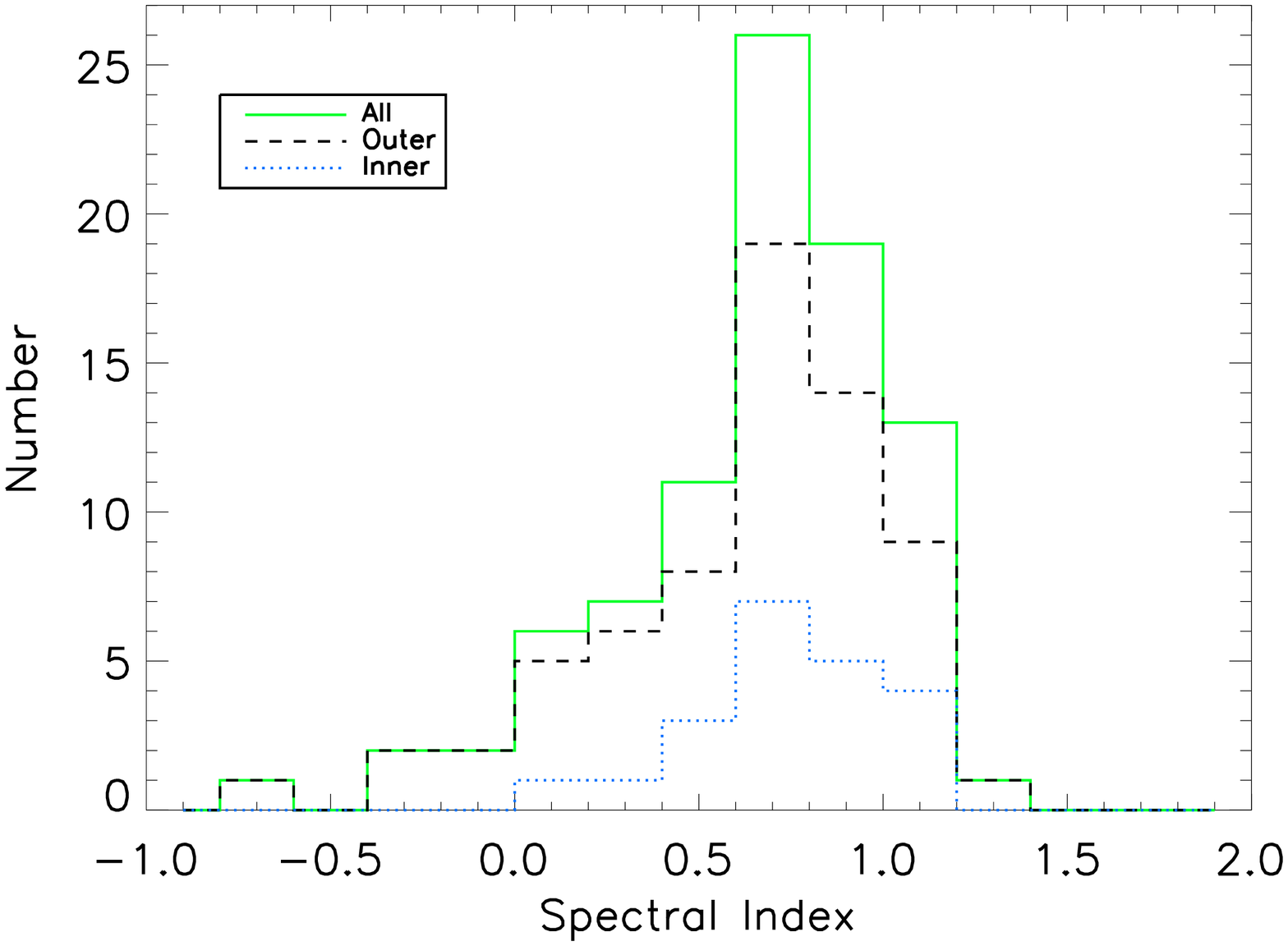}
\caption{Histogram of the spectral index distribution for radio sources in cluster fields.
The spectral index $\alpha$ is defined by $S \sim \nu^{-\alpha}$.
The distribution for 88 sources is shown with a solid (green) line, the
distribution for the 67 sources in the outer regions of cluster fields ($r > 0.5$ arcmin)
is shown with a dashed (black) line, and the distribution for the 21 sources in the inner
regions of cluster fields ($r < 0.5$ arcmin) is shown with a dotted (blue) line.
The overall mean spectral index is $\alpha = 0.66$ with an rms dispersion of $0.36$.
The mean spectral index for the outer regions of cluster fields
is $\alpha = 0.63$ with an rms dispersion of $0.38$ and the mean spectral index
for the the inner regions of cluster fields is $\alpha = 0.75$ with an rms dispersion of $0.24$.
The medians are $0.72$, $0.71$, and $0.76$ for all, outer, and inner, respectively.}
\label{fig1}
\end{figure}


\clearpage

\begin{figure}
\epsscale{1.0}
\plotone{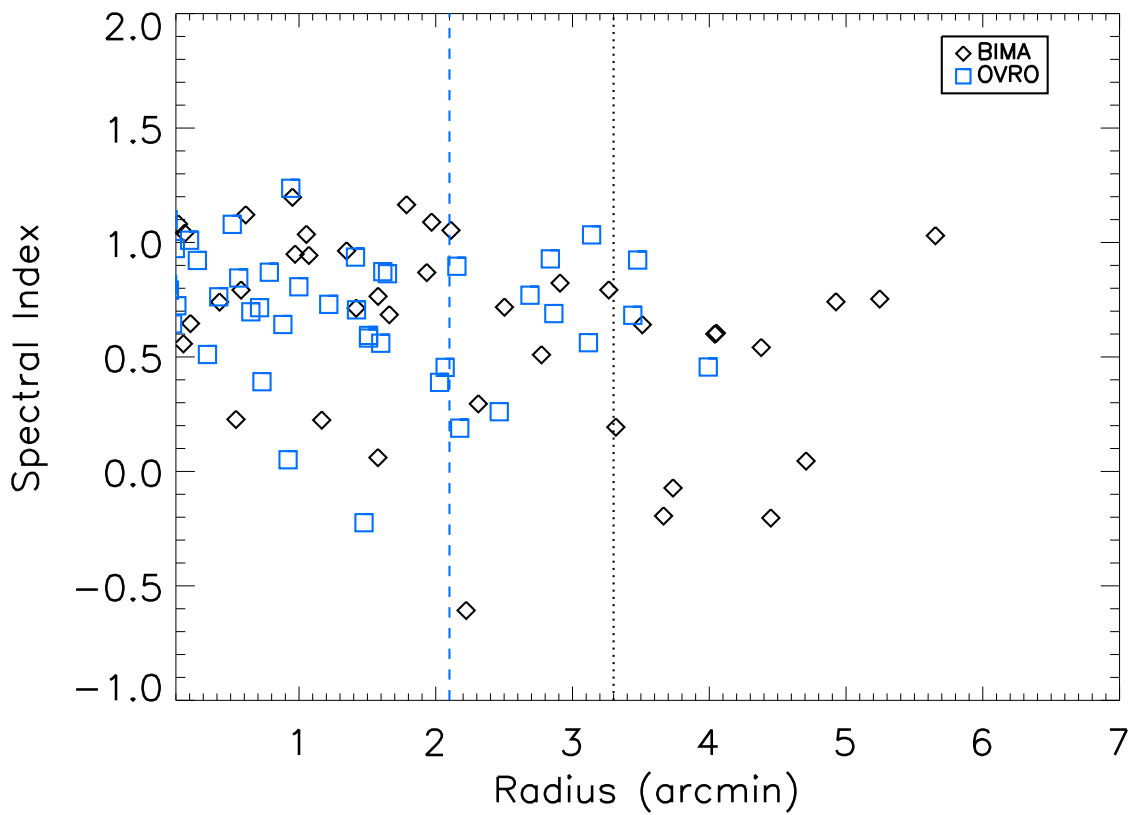}
\caption{Spectral index as a function of radius for 88 cluster sources.
The figure shows no clear trend with radius.
The half-power point for BIMA is shown as a dotted (black) line and the half-power point
for OVRO is shown as a dashed (blue) line.
Beyond the half power point, the beam attenuation factor becomes a
potentially important source of systematic uncertainty.
When a source has observations from both BIMA and OVRO, we
choose the field with the best combination of sensitivity and survey
area; repeat sources are not shown here. }
\label{fig2}
\end{figure}


\clearpage

\begin{figure}
\epsscale{1.0}
\plotone{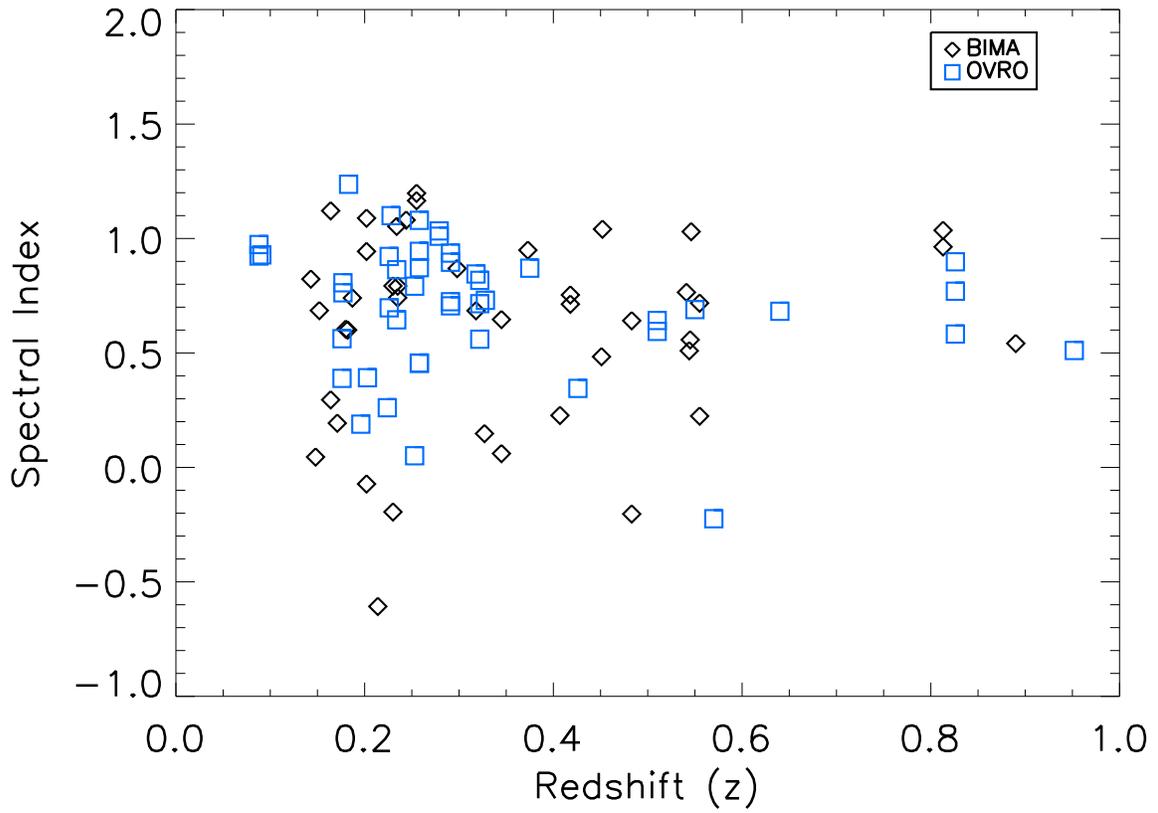}
\caption{Spectral index as a function of redshift for 88 cluster sources.
The figure shows no clear trend with redshift.
When a source has observations from both BIMA and OVRO, we
choose the field with the best combination of sensitivity and survey
area; repeat sources are not shown here. }
\label{fig3}
\end{figure}


\clearpage

\begin{figure}
\epsscale{1.0}
\plotone{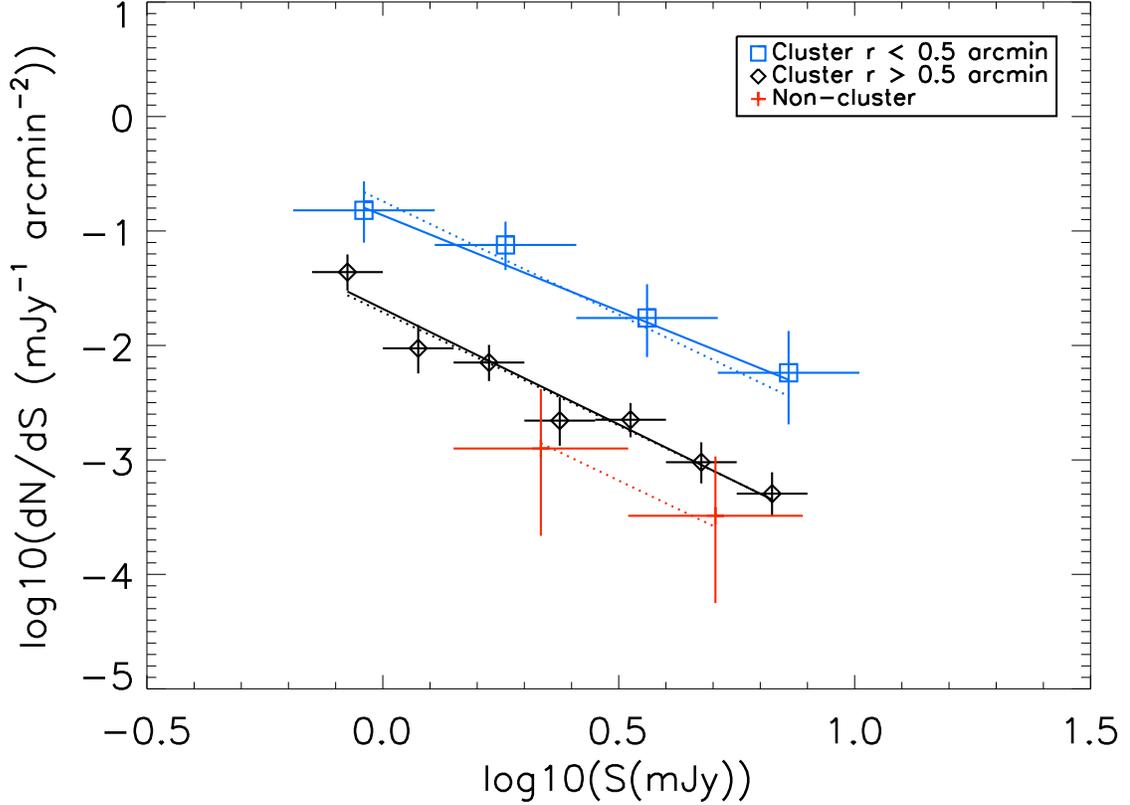}
\caption{Differential number counts, $log_{10}(dN/dS)$, as a function of flux, $log_{10}(S)$,
and best fit power laws for the central regions of cluster fields ($r < 0.5$ arcmin),
outer regions of cluster fields ($r > 0.5$ arcmin), and non-cluster fields.
Solid lines indicate the best fit power laws for each set individually
and dotted lines indicate the best fits using a Markov chain algorithm to
simultaneously estimate the normalizations and a common power law index.
Using the best joint fit normalizations,
we find that counts toward the outer regions of clusters are a factor
of $3.3^{+4.1}_{-1.8}$ higher than counts in the field.
Counts toward the inner regions of clusters are a factor of $8.9^{+4.3}_{-2.8}$ higher than the outer regions.
The outer boundary used for the outer regions of cluster fields is set by the noise levels
in the fields as described in Section 4.1.
Error bars on the data come from Poisson errors on raw counts
and do not include other sources of uncertainty.}
\label{fig4}
\end{figure}


\clearpage

\begin{figure}
\epsscale{1.0}
\plotone{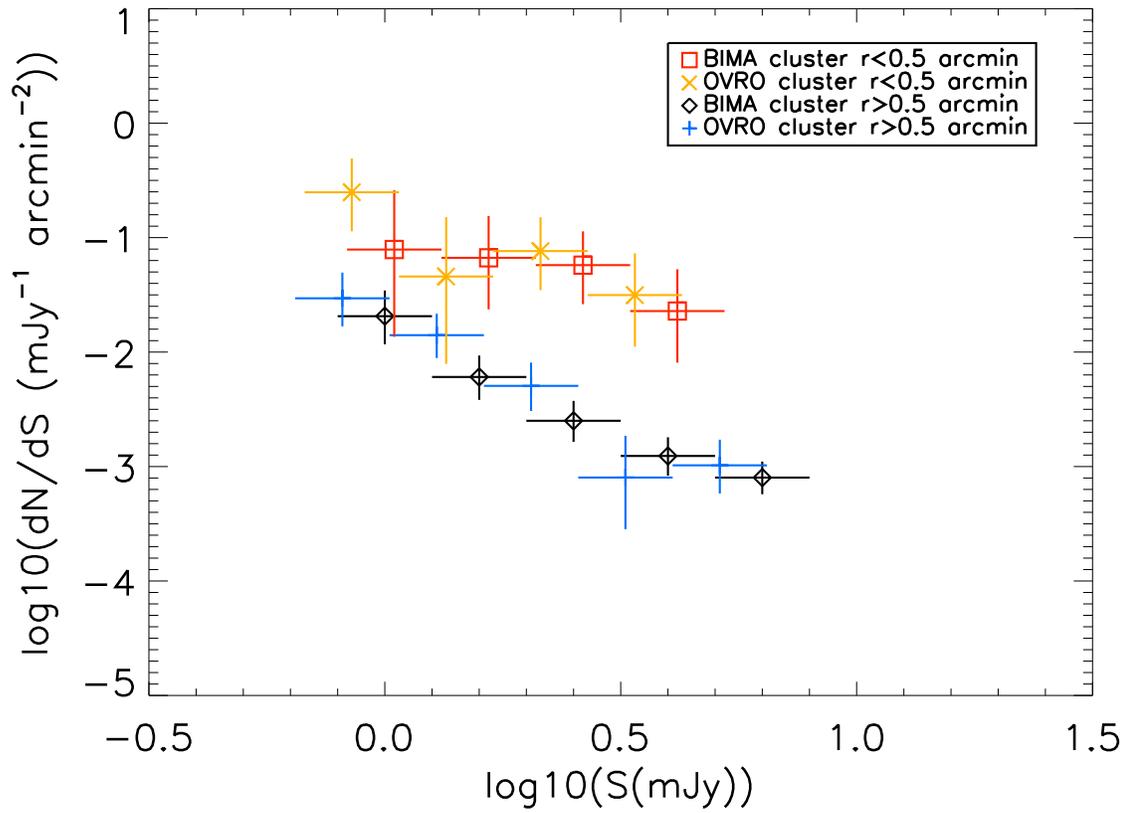}
\caption{Differential number counts as a function of flux
for cluster fields taken at BIMA and OVRO. They are in good agreement.}
\label{fig5}
\end{figure}


\clearpage

\begin{figure}
\epsscale{1.0}
\plotone{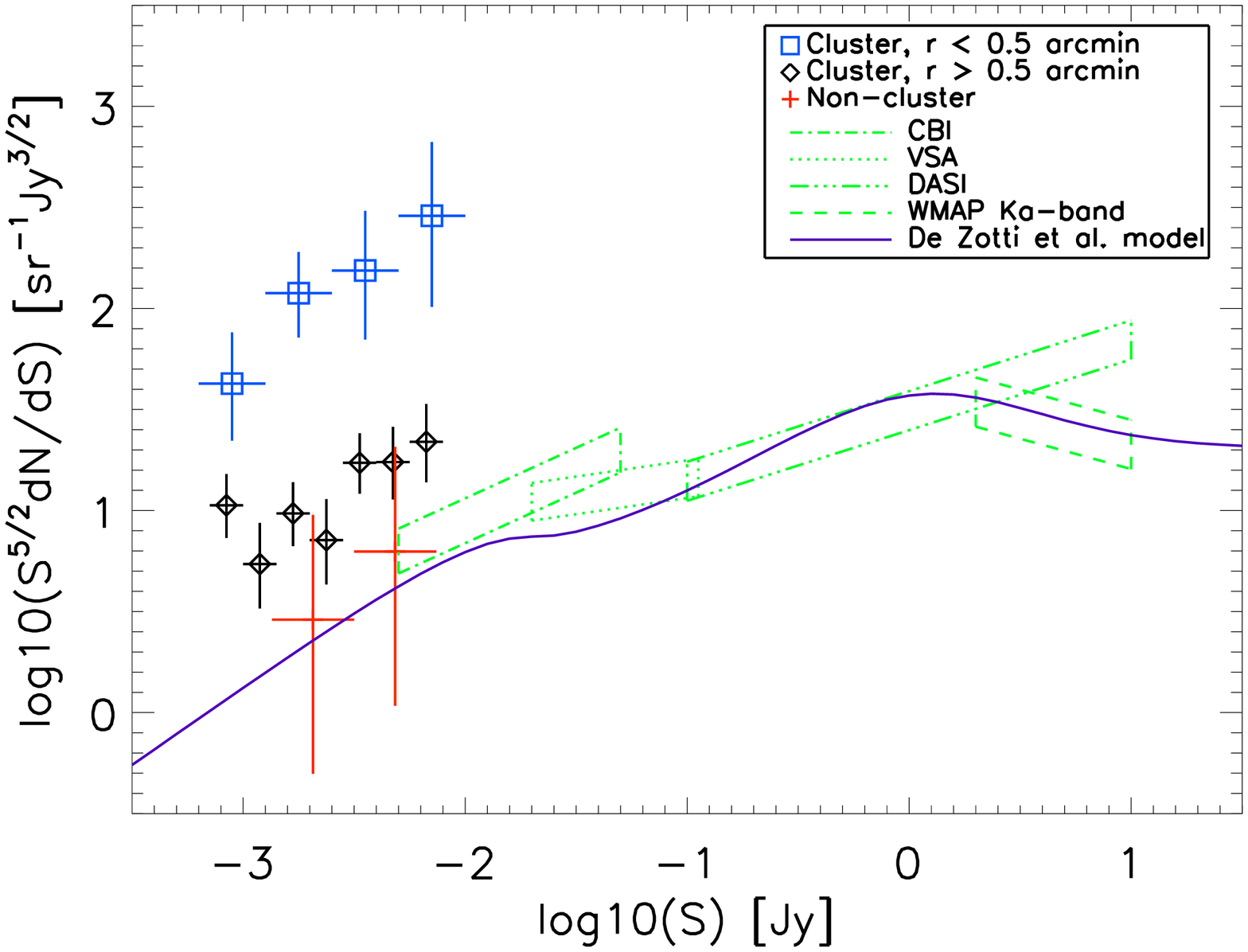}
\caption{Differential number counts as a function of flux.
Sources are overdense in the cluster fields, especially within the central 
arcminute.
Our SZE data are shown along with measurements from CBI \citep{mason2003},
VSA \citep{cleary2005}, DASI \citep{kovac2002}, and WMAP Ka-band \citep{bennett2003}
as well as the 30 GHz model from \citet{dezotti2005}.
Counts in the non-cluster fields are consistent with the model and extrapolations from other experiments.}
\label{fig6}
\end{figure}




\clearpage
\begin{deluxetable}{llcrrrrrrrrrcccc}
\tablecolumns{16}
\rotate
\tabletypesize{\tiny}
\tablewidth{0pt}
\tablecaption{Radio Sources in Cluster Fields \label{tbl1}}
\tablehead{
\colhead{} &
\colhead{} &
\colhead{} &
\multicolumn{2}{c}{Pointing Center} &
\colhead{} &
\colhead{28.5 GHz} &
\multicolumn{2}{c}{SZ} & \multicolumn{6}{c}{28.5 GHz Source} &
\colhead{1.4GHz}
\\
\colhead{} &
\colhead{} &
\colhead{} &
\multicolumn{2}{c}{\hrulefill} &
\colhead{} &
\colhead{RMS} &
\multicolumn{2}{c}{\hrulefill} &
\multicolumn{6}{c}{\hrulefill} &
\colhead{Flux}
\\
\colhead{Field} &
\colhead{z} &
\colhead{ref} &
\colhead{$\alpha$(J2000)} &
\colhead{$\delta$(J2000)} &
\colhead{Array} &
\colhead{(mJy)} &
\colhead{$\Delta\alpha('')$} &
\colhead{$\Delta\delta('')$} &
\colhead{src} &
\colhead{$\alpha$(J2000)} &
\colhead{$\delta$(J2000)} &
\colhead{Radius ('')} &
\colhead{Flux (mJy)} &
\colhead{Unc. (mJy)} &
\colhead{(mJy)}
}
\startdata
CL~0016+16 & 0.546 & \ref{S91} & 00:18:34.6 & +16:26:18.2 &  B & 0.131 & -21 & -14 &  1 &00:18:31.2 &+16:20:42.6 & 322.7 &  12.06 & 1.22 & 268.8 \\
 & & & & &  O & 0.074 &  &  &   & \nd & \nd & \nd & \nd & \nd & \nd \\
CL~J0018.8+1602 & 0.541 & \ref{H98} & 00:18:47.9 & +16:02:22.0 &  B & 0.180 &\nd & \nd &  1 &00:18:41.4 &+16:02:08.5 & 94.7 &  2.65 & 0.21 & 26.6 \\
MACS~J0025$-$12 & 0.586 & \ref{0025} & 00:22:57.8 & -12:39:06.6 &  B & 0.312 &\nd & \nd &   & \nd & \nd & \nd & \nd & \nd & \nd \\
 & & & & &  O & 0.236 &  &  &  1 &00:22:59.9 &-12:39:40.2 & 46.0 &  1.43 & 0.26 & \tblntmrk{a} \\
Cl~0024+1654 & 0.393 & \ref{DRESS} & 00:26:35.8 & +17:09:41.0 &  B & 0.085 &\nd & \nd &   & \nd & \nd & \nd & \nd & \nd & \nd \\
Abell~68 & 0.255 & \ref{S99} & 00:37:06.6 & +09:09:18.6 &  B & 0.097 & -17 & 30 &  1 &00:37:07.6 &+09:08:23.4 & 90.4 &  1.60 & 0.10 & 59.1 \\
 & & & & &  B & 0.097 &  &  &  2 &00:37:07.0 &+09:07:58.7 & 111.9 &  1.38 & 0.11 & \tblntmrk{b}  \\
 & & & & &  B & 0.097 &  &  &  3 &00:37:06.6 &+09:07:31.4 & 137.7 &  1.20 & 0.12 & 40.2 \\
MACS~J0111+08 & \nd & \nd & 01:11:34.3 & +08:55:53.0 &  B & 0.133 &\nd & \nd &   & \nd & \nd & \nd & \nd & \nd & \nd \\
Abell~267 & 0.230 & \ref{S99} & 01:52:41.9 & +01:00:24.1 &  B & 0.101 &  5 &  2 &  1 &01:52:54.7 &+01:02:11.4 & 214.5 &  7.55 & 0.24 & 4.2 \\
 & & & & &  B & 0.101 &  &  &  2 &01:52:29.2 &+00:59:38.5 & 201.4 &  2.75 & 0.20 & 30.0 \\
 & & & & &  O & 0.073 &  &  &  1 &01:52:54.6 &+01:02:10.6 & 212.4 &  5.53 & 0.73 & 4.2 \\
CL~J0152.7$-$1357 & 0.830 & \ref{R00} & 01:52:43.0 & -13:57:29.0 &  B & 0.183 & -1 & -9 &   & \nd & \nd & \nd & \nd & \nd & \nd \\
Abell~348 & 0.274 & \ref{S99} & 02:23:59.0 & -08:35:39.4 &  O & 0.102 &\nd & \nd &   & \nd & \nd & \nd & \nd & \nd & \nd \\
RCS~J0224.5$-$0002 & 0.773 & \ref{G03} & 02:24:34.1 & -00:02:30.9 &  B & 0.112 & -17 & -1 &   & \nd & \nd & \nd & \nd & \nd & \nd \\
 & & & & &  O & 0.094 &  &  &   & \nd & \nd & \nd & \nd & \nd & \nd \\
Abell~370 & 0.375 & \ref{S99} & 02:39:52.5 & -01:34:20.1 &  B & 0.152 & -2 & -24 &  1 &02:39:55.5 &-01:34:05.6 & 60.3 &  1.18 & 0.16 & 10.5 \\
 & & & & &  O & 0.072 &  &  &  1 &02:39:55.5 &-01:34:06.7 & 59.5 &  0.76 & 0.08 & 10.5 \\
Abell~383 & 0.187 & \ref{S99} & 02:48:03.6 & -03:32:09.6 &  B & 0.465 &\nd & \nd &  1 &02:48:03.4 &-03:31:44.6 & 25.2 &  4.40 & 0.47 & 40.9 \\
MS~0302.7+1658 & 0.426 & \ref{G94} & 03:05:31.7 & +17:10:02.8 &  B & 0.267 &\nd & \nd &  1 &03:05:31.5 &+17:10:03.7 & 2.7 &  2.70 & 0.27 & 4.8 \\
 & & & & &  O & 0.047 &  &  &  1 &03:05:31.7 &+17:10:02.4 & 0.4 &  1.70 & 0.05 & 4.8 \\
MACS~J0329$-$02 & 0.467 & \ref{0329} & 03:29:40.5 & -02:11:40.0 &  O & 0.165 &\nd & \nd &   & \nd & \nd & \nd & \nd & \nd & \nd \\
Abell~478 & 0.088 & \ref{S99} & 04:13:26.2 & +10:27:57.6 &  B & 0.116 & -6 & -9 &  1 &04:13:25.3 &+10:27:54.8 & 9.9 &  2.43 & 0.12 & 36.9 \\
 & & & & &  B & 0.116 &  &  &  2 &04:13:38.3 &+10:28:06.9 & 185.0 &  2.17 & 0.20 & 47.7 \\
 & & & & &  O & 0.059 &  &  &  1 &04:13:26.5 &+10:27:53.8 & 11.5 &  1.96 & 0.06 & 36.9 \\
 & & & & &  O & 0.059 &  &  &  2 &04:13:40.3 &+10:28:05.3 & 215.0 &  2.95 & 0.47 & 47.7 \\
RX~J0439.0+0715 & 0.244 & \ref{E98} & 04:39:01.2 & +07:15:36.0 &  B & 0.207 &\nd & \nd &  1 &04:39:01.2 &+07:15:28.9 & 7.1 &  1.18 & 0.21 & 30.6 \\
MS~0451.6$-$0305 & 0.550 & \ref{G94} & 04:54:10.8 & -03:00:56.8 &  O & 0.050 & 11 &  4 &  1 &04:54:22.1 &-03:01:25.0 & 161.6 &  1.80 & 0.19 & 14.4 \\
Abell~520 & 0.202 & \ref{S99} & 04:54:12.7 & +02:55:23.9 &  B & 0.103 & -42 & -44 &  1 &04:54:01.1 &+02:57:46.7 & 228.3 &  7.83 & 0.25 & 6.3 \\
 & & & & &  B & 0.103 &  &  &  2 &04:54:16.9 &+02:55:32.3 & 118.4 &  0.84 & 0.11 & 14.4 \\
 & & & & &  B & 0.103 &  &  &  3 &04:54:20.3 &+02:54:56.2 & 158.0 &  1.00 & 0.13 & 26.5 \\
 & & & & &  O & 0.078 &  &  &  1 &04:54:01.3 &+02:57:46.3 & 226.6 &  4.42 & 0.88 & 6.3 \\
 & & & & &  O & 0.078 &  &  &  2 &04:54:16.8 &+02:55:31.9 & 116.2 &  1.09 & 0.09 & 14.4 \\
 & & & & &  O & 0.078 &  &  &  3 &04:54:20.4 &+02:54:56.8 & 159.1 &  0.74 & 0.14 & 26.5 \\
MACS~J0647.7+7015 & 0.584 & \ref{L03} & 06:47:50.0 & +70:14:55.0 &  O & 0.061 &  3 &  1 &   & \nd & \nd & \nd & \nd & \nd & \nd \\
MACS~J0717.5+3745 & 0.555 & \ref{E03} & 07:17:33.8 & +37:45:20.0 &  B & 0.172 & -42 &  0 &  1 &07:17:37.2 &+37:44:22.7 & 100.4 &  3.29 & 0.19 & 6.5\tblntmrk{c}  \\
 & & & & &  B & 0.172 &  &  &  2 &07:17:41.1 &+37:43:17.5 & 178.0 &  2.28 & 0.25 & 19.8 \\
Abell~586 & 0.171 & \ref{S99} & 07:32:20.3 & +31:38:02.0 &  B & 0.113 & -10 & -7 &  1 &07:32:20.3 &+31:41:21.0 & 205.9 &  4.42 & 0.23 & 7.9\tblntmrk{c} \\
MS~0735.6+7421 & 0.216 & \ref{G94} & 07:41:45.0 & +74:14:36.7 &  O & 0.209 &\nd & \nd &   & \nd & \nd & \nd & \nd & \nd & \nd \\
MACS~J0744.8+3937 & 0.686 & \ref{L03} & 07:44:52.5 & +39:27:30.0 &  B & 0.241 & -2 &  3 &  & \nd & \nd & \nd & \nd & \nd & \nd \\
 & & & & &  O & 0.069 &  &  &   & \nd & \nd & \nd & \nd & \nd & \nd \\
Abell~611 & 0.288 & \ref{S99} & 08:00:56.7 & +36:03:21.7 &  O & 0.041 & -3 &  1 &   & \nd & \nd & \nd & \nd & \nd & \nd \\
Abell~665 & 0.182 & \ref{S99} & 08:30:59.3 & +65:50:09.5 &  B & 0.101 & -11 & 40 &  1 &08:31:30.8 &+65:52:36.2 & 229.8 &  4.95 & 0.29 & 30.2 \\
 & & & & &  O & 0.076 &  &  &   & \nd & \nd & \nd & \nd & \nd & \nd \\
Abell~697 & 0.282 & \ref{S99} & 08:42:57.6 & +36:21:59.4 &  O & 0.041 &  3 & -5 &   & \nd & \nd & \nd & \nd & \nd & \nd \\
Cl~0847.2+3617 & 0.373 & \ref{C99} & 08:50:10.1 & +36:05:09.6 &  B & 0.160 &\nd & \nd &  1 &08:50:13.0 &+36:04:22.9 & 58.3 &  1.19 & 0.17 & 20.8 \\
Zw~2089 & 0.235 & \ref{D02} & 09:00:37.9 & +20:54:57.6 &  B & 0.232 &\nd & \nd &   & \nd & \nd & \nd & \nd & \nd & \nd \\
Abell~750 & 0.180 & \ref{S99} & 09:09:11.8 & +10:59:20.4 &  B & 0.160 &\nd & \nd &  1 &09:09:03.4 &+11:02:50.3 & 243.1 &  3.69 & 0.46 & 22.8 \\
MACS~J0913+40 & 0.452 & \ref{0913} & 09:13:46.0 & +40:56:20.0 &  B & 0.117 &  2 &  7 &  1 &09:13:45.4 &+40:56:26.7 & 9.5 &  0.69 & 0.12 & 15.9 \\
 & & & & &  O & 0.236 &  &  &   & \nd & \nd & \nd & \nd & \nd & \nd \\
Abell~773 & 0.217 & \ref{S99} & 09:17:54.5 & +51:43:43.5 &  B & 0.134 & -15 &  6 &   & \nd & \nd & \nd & \nd & \nd & \nd \\
 & & & & &  O & 0.078 &  &  &   & \nd & \nd & \nd &  \nd & \nd & \nd \\
Abell~781 & 0.298 & \ref{S99} & 09:20:28.8 & +30:31:08.4 &  B & 0.235 &\nd & \nd &  1 &09:20:22.6 &+30:29:44.5 & 116.1 &  5.33 & 0.30 & 73.1 \\
Abell~851 & 0.407 & \ref{S99} & 09:42:56.6 & +46:59:20.4 &  B & 0.149 &  1 & -10 &  1 &09:42:57.5 &+46:58:49.2 & 22.3 &  1.06 & 0.15 & 2.1\tblntmrk{c} \\
Zwicky~2701 & 0.214 & \ref{C95} & 09:52:47.5 & +51:53:27.6 &  B & 0.474 &\nd & \nd &  1 &09:52:43.1 &+51:51:20.6 & 133.3 &  18.70 & 0.64 & 3.0 \\
Abell~959 & 0.353 & \ref{S99} & 10:17:35.9 & +59:34:05.6 &  O & 0.089 &\nd & \nd &   & \nd & \nd & \nd & \nd & \nd & \nd \\
Zwicky~3146 & 0.291 & \ref{A92} & 10:23:39.7 & +04:11:11.0 &  B & 0.163 &  5 &  6 &  1 &10:23:45.0 &+04:10:40.8 & 82.4 &  5.35 & 0.18 & 95.8 \\
 & & & & &  B & 0.163 &  &  &  2 &10:23:37.2 &+04:09:04.5 & 139.4 &  2.03 & 0.22 & 31.5 \\
 & & & & &  O & 0.074 &  &  &  1 &10:23:45.0 &+04:10:40.3 & 82.7 &  5.70 & 0.10 & 95.8 \\
 & & & & &  O & 0.074 &  &  &  2 &10:23:37.1 &+04:09:08.0 & 136.7 &  2.12 & 0.15 & 31.5 \\
 & & & & &  O & 0.074 &  &  &  3 &10:23:44.9 &+04:11:45.4 & 78.2 &  0.85 & 0.10 & 7.1 \\
 & & & & &  O & 0.074 &  &  &  4 &10:23:39.3 &+04:11:11.5 & 12.8 &  0.41 & 0.07 & 3.6 \\
Abell~992 & 0.247 & \ref{S99} & 10:22:33.7 & +20:29:29.8 &  O & 0.071 &\nd & \nd &   & \nd & \nd & \nd & \nd & \nd & \nd \\
Abell~990 & 0.144 & \ref{S99} & 10:23:39.8 & +49:08:38.5 &  B & 0.852 &\nd & \nd &   & \nd & \nd & \nd & \nd & \nd & \nd \\
MS~1054.4$-$0321 & 0.826 & \ref{L95} & 10:56:59.5 & -03:37:28.2 &  O & 0.060 & -5 & -8 &  1 &10:56:59.5 &-03:37:26.9 & 11.0 &  0.94 & 0.06 & 14.1 \\
 & & & & &  O & 0.060 &  &  &  2 &10:56:48.8 &-03:37:26.5 & 156.3 &  1.79 & 0.19 & 18.2 \\
 & & & & &  O & 0.060 &  &  &  3 &10:56:57.9 &-03:38:55.2 & 81.2 &  0.54 & 0.08 & 3.1 \\
MACS~J1108+09 & 0.480 & \ref{1108} & 11:08:55.5 & +09:06:00.0 &  O & 0.113 &\nd & \nd &   & \nd & \nd & \nd & \nd & \nd & \nd \\
MACS~J1115+53 & 0.510 & \ref{1115} & 11:15:14.9 & +53:19:56.0 &  O & 0.077 & -4 & 14 &  1 &11:15:17.2 &+53:19:07.3 & 67.1 &  1.06 & 0.09 & 7.3\tblntmrk{c} \\
 & & & & &  O & 0.077 &  &  &  2 &11:15:21.3 &+53:18:46.0 & 103.7 &  0.75 & 0.11 & 4.5\tblntmrk{c} \\
Zwicky~5247 & 0.229 & \ref{B00} & 12:34:17.3 & +09:46:12.0 &  B & 0.379 &\nd & \nd &   & \nd & \nd & \nd & \nd & \nd & \nd \\
MS~1137+66 & 0.782 & \ref{D99} & 11:40:23.9 & +66:08:19.1 &  B & 0.081 & -12 & -14 &   & \nd & \nd & \nd & \nd & \nd & \nd \\
Abell~1351 & 0.322 & \ref{S99} & 11:42:24.6 & +58:32:06.5 &  O & 0.122 &\nd & \nd &  1 &11:42:24.4 &+58:32:04.5 & 2.4 &  6.37 & 0.12 & 74.6\tblntmrk{c} \\
 & & & & &  O & 0.122 &  &  &  2 &11:42:13.7 &+58:31:23.3 & 95.9 &  1.86 & 0.18 & 10.0\tblntmrk{c} \\
 & & & & &  O & 0.122 &  &  &  3 &11:42:22.9 &+58:31:25.9 & 42.7 &  0.84 & 0.13 & 7.2\tblntmrk{c} \\
MACS~J1149.5+2223 & 0.544 & \ref{L03} & 11:49:34.3 & +22:23:42.5 &  B & 0.113 & 10 & 12 &  1 &11:49:22.3 &+22:23:29.2 & 177.4 &  3.40 & 0.18 & 15.8 \\
 & & & & &  O & 0.243 &  &  &   & \nd & \nd & \nd & \nd & \nd & \nd \\
Abell~1413 & 0.143 & \ref{S99} & 11:55:18.0 & +23:24:18.9 &  B & 0.146 & -4 & 21 &  1 &11:55:08.7 &+23:26:17.0 & 158.0 &  2.36 & 0.25 & 28.1 \\
CL~1226+33 & 0.890 & \ref{E01} & 12:26:58.0 & +33:32:45.0 &  B & 0.116 &  1 & 13 &  1 &12:27:18.8 &+33:32:06.0 & 264.4 &  5.83 & 0.41 & 29.8 \\
Abell~1576 & 0.279 & \ref{S99} & 12:36:59.3 & +63:11:10.3 &  O & 0.102 &\nd & \nd &  1 &12:36:57.6 &+63:11:11.7 & 12.0 &  0.88 & 0.10 & 18.4 \\
 & & & & &  O & 0.102 &  &  &  2 &12:36:32.4 &+63:11:58.6 & 188.4 &  2.93 & 0.52 & 65.9 \\
Abell~1682 & 0.234 & \ref{S99} & 13:06:57.2 & +46:32:42.0 &  B & 0.422 &\nd & \nd &  1 &13:06:45.9 &+46:33:30.7 & 126.7 &  8.10 & 0.55 & 193.6 \\
MACS~J1311.0$-$0311 & 0.519 & \ref{1311} & 13:11:01.7 & -03:10:39.0 &  B & 0.123 &  9 & -6 &   & \nd & \nd & \nd & \nd & \nd & \nd \\
Abell~1689 & 0.183 & \ref{S99} & 13:11:30.3 & -01:20:25.4 &  B & 0.169 & -18 & -4 &  1 &13:11:31.2 &-01:19:34.5 & 63.2 &  1.30 & 0.18 & 59.6 \\
 & & & & &  O & 0.059 &  &  &  1 &13:11:31.5 &-01:19:32.0 & 67.9 &  1.43 & 0.07 & 59.6 \\
Abell~1703 & 0.258 & \ref{S99} & 13:15:05.3 & +51:49:01.9 &  O & 0.243 &\nd & \nd &  1 &13:15:08.5 &+51:48:53.7 & 30.7 &  1.74 & 0.25 & 45.0 \\
Abell~1704 & 0.221 & \ref{S99} & 13:14:26.0 & +64:34:41.2 &  O & 0.098 &\nd & \nd &   & \nd & \nd & \nd & \nd & \nd & \nd \\
Abell~1722 & 0.328 & \ref{S99} & 13:20:09.1 & +70:04:38.6 &  O & 0.124 &\nd & \nd &  1 &13:20:14.6 &+70:05:46.2 & 73.0 &  0.93 & 0.15 & 8.4 \\
RCS~J1324.5+2844 & 0.997 & \ref{G05} & 13:24:28.3 & +28:44:58.5 &  O & 0.065 &\nd & \nd &   & \nd & \nd & \nd & \nd & \nd & \nd \\
RCS~J1326.5+2903 & 0.952 & \ref{G05} & 13:26:31.1 & +29:03:19.8 &  B & 0.140 & -4 & -15 &   & \nd & \nd & \nd & \nd & \nd & \nd \\
 & & & & &  O & 0.065 &  &  &  1 &13:26:31.9 &+29:03:36.5 & 34.8 &  0.60 & 0.07 & 2.8 \\
Abell~1763 & 0.228 & \ref{S99} & 13:35:20.2 & +41:00:04.0 &  O & 0.407 &\nd & \nd &  1 &13:35:20.0 &+41:00:02.3 & 2.6 &  31.20 & 0.41 & 857.2 \\
RX~J1347.5$-$1145 & 0.451 & \ref{S95} & 13:47:30.7 & -11:45:08.6 &  B & 0.188 & -2 &  4 &  1 &13:47:30.6 &-11:45:09.2 & 4.5 &  10.68 & 0.19 & 45.9 \\
 & & & & &  O & 0.235 &  &  &  1 &13:47:30.6 &-11:45:07.7 & 2.9 &  9.93 & 0.23 & 45.9 \\
MS~1358.4+6245 & 0.327 & \ref{G94} & 13:59:50.6 & +62:31:05.3 &  B & 0.084 & -6 &  2 &  1 &13:59:50.7 &+62:31:05.6 & 6.6 &  1.67 & 0.08 & 2.6\tblntmrk{c} \\
 & & & & &  O & 0.089 &  &  &  1 &13:59:50.6 &+62:31:04.6 & 6.5 &  1.51 & 0.09 & 2.6\tblntmrk{c} \\
Abell~1835 & 0.253 & \ref{S99} & 14:01:02.0 & +02:52:41.7 &  B & 0.137 & -3 &  4 &  1 &14:01:02.1 &+02:52:43.0 & 4.3 &  3.31 & 0.14 & 31.2\tblntmrk{c} \\
 & & & & &  B & 0.137 &  &  &  2 &14:01:00.4 &+02:51:51.0 & 58.6 &  1.26 & 0.14 & 1.6\tblntmrk{c} \\
 & & & & &  O & 0.073 &  &  &  1 &14:01:02.1 &+02:52:44.5 & 4.6 &  2.88 & 0.07 & 31.2\tblntmrk{c} \\
 & & & & &  O & 0.073 &  &  &  2 &14:01:00.5 &+02:51:51.2 & 57.7 &  1.36 & 0.08 & 1.6\tblntmrk{c} \\
RCS~J1419.2+5326 & 0.640 & \ref{G03} & 14:19:12.1 & +53:26:11.4 &  O & 0.117 &\nd & \nd &  1 &14:19:24.2 &+53:23:15.5 & 206.4 &  15.66 & 0.89 & 122.3 \\
MACS~J1423.8+2404 & 0.545 & \ref{L03} & 14:23:48.3 & +24:04:47.5 &  B & 0.121 & -9 & -10 &  1 &14:23:47.7 &+24:04:43.1 & 5.9 &  1.49 & 0.12 & 8.0 \\
Abell~1914 & 0.171 & \ref{S99} & 14:26:03.5 & +37:49:46.5 &  B & 0.150 & -32 & -8 &   & \nd & \nd & \nd & \nd & \nd & \nd \\
Abell~1995 & 0.318 & \ref{P00} & 14:52:57.6 & +58:02:55.7 &  B & 0.096 &  7 &  1 &  1 &14:52:47.5 &+58:01:56.1 & 106.1 &  0.62 & 0.11 & 4.9 \\
 & & & & &  O & 0.051 &  &  &  2 &14:53:00.6 &+58:03:19.4 & 27.9 &  0.60 & 0.05 & 7.7 \\
MS~1455.0+2232 & 0.258 & \ref{G94} & 14:57:15.1 & +22:20:34.3 &  B & 0.321 &  &  &  1 &14:56:59.3 &+22:18:59.2 & 239.0 &  6.35 & 0.89 & 19.3 \\
 & & & & &  O & 0.037 &  &  &  1 &14:56:59.2 &+22:19:01.1 & 239.5 &  4.88 & 0.69 & 19.3 \\
 & & & & &  O & 0.037 &  &  &  2 &14:57:15.1 &+22:20:34.8 & 0.7 &  0.96 & 0.04 & 16.5 \\
 & & & & &  O & 0.037 &  &  &  3 &14:57:08.3 &+22:20:14.7 & 96.8 &  0.95 & 0.05 & 13.2 \\
 & & & & &  O & 0.037 &  &  &  4 &14:57:10.8 &+22:18:45.6 & 124.1 &  0.99 & 0.07 & 3.9 \\
RX~J1532.9+3021 & 0.345 & \ref{D02} & 15:32:54.2 & +30:21:10.8 &  B & 0.176 & 34 & -8 &  1 &15:32:50.9 &+30:19:46.2 & 108.3 &  6.58 & 0.20 & 7.9 \\
 & & & & &  B & 0.176 &  &  &  2 &15:32:53.8 &+30:20:59.4 & 39.2 &  3.25 & 0.18 & 22.8 \\
Abell~2111 & 0.229 & \ref{S99} & 15:39:41.8 & +34:25:01.2 &  B & 0.091 & -23 &  1 &   & \nd & \nd & \nd & \nd & \nd & \nd \\
Abell~2142 & 0.091 & \ref{S99} & 15:58:20.2 & +27:13:52.0 &  B & 0.159 & 17 & 20 &  1 &15:58:14.1 &+27:16:20.2 & 161.9 &  7.81 & 0.26 & 107.3 \\
 & & & & &  O & 0.063 &  &  &  1 &15:58:14.1 &+27:16:21.5 & 163.2 &  6.54 & 0.23 & 107.3 \\
Abell~2146 & 0.234 & \ref{S99} & 15:56:14.4 & +66:20:56.2 &  O & 0.148 &\nd & \nd &  1 &15:56:13.8 &+66:20:53.5 & 4.5 &  2.19 & 0.15 & 15.3 \\
 & & & & &  O & 0.148 &  &  &  2 &15:56:04.2 &+66:22:13.9 & 98.8 &  3.01 & 0.22 & 40.6 \\
 & & & & &  O & 0.148 &  &  &  3 &15:55:58.0 &+66:20:04.4 & 111.4 &  1.43 & 0.25 & \tblntmrk{d} \\
Abell~2163 & 0.203 & \ref{S99} & 16:15:46.0 & -06:08:55.0 &  B & 0.169 & 38 &  8 &  1 &16:15:43.6 &-06:08:42.3 & 74.5 &  1.31 & 0.17 & 4.8\tblntmrk{e} \\
 & & & & &  O & 0.071 &  &  &  1 &16:15:43.3 &-06:08:40.5 & 79.0 &  1.47 & 0.08 & 4.8\tblntmrk{e} \\
RCS~J1620.2+2929 & 0.870 & \ref{G03} & 16:20:10.0 & +29:29:21.5 &  O & 0.106 &\nd & \nd &   & \nd & \nd & \nd & \nd & \nd & \nd \\
MACS~J1621.3+3810 & 0.465 & \ref{E03} & 16:21:24.0 & +28:10:02.0 &  B & 0.127 & 26 &  1 &   & \nd & \nd & \nd & \nd & \nd & \nd \\
 & & & & &  O & 0.089 &  &  &   & \nd & \nd & \nd & \nd & \nd & \nd \\
Abell~2204 & 0.152 & \ref{S99} & 16:32:46.9 & +05:34:32.4 &  B & 0.129 & -7 & -11 &  1 &16:32:47.0 &+05:34:33.2 & 14.4 &  8.79 & 0.13 & 69.3 \\
Abell~2218 & 0.176 & \ref{S99} & 16:35:49.5 & +66:12:44.4 &  B & 0.124 & 12 & -16 &  1 &16:35:22.3 &+66:13:20.4 & 183.8 &  5.17 & 0.20 & \tblntmrk{f} \\
 & & & & &  B & 0.124 &  &  &  2 &16:36:15.0 &+66:14:23.6 & 183.2 &  3.09 & 0.22 & 12.1\tblntmrk{e} \\
 & & & & &  B & 0.124 &  &  &  3 &16:35:47.6 &+66:14:44.6 & 138.4 &  1.41 & 0.16 & 4.8 \tblntmrk{e}\\
 & & & & &  O & 0.054 &  &  &  1 &16:35:22.1 &+66:13:22.2 & 185.6 &  4.45 & 0.20 & \tblntmrk{f} \\
 & & & & &  O & 0.054 &  &  &  2 &16:36:15.8 &+66:14:22.3 & 186.5 &  2.22 & 0.27 & 12.1\tblntmrk{e} \\
 & & & & &  O & 0.054 &  &  &  3 &16:35:47.7 &+66:14:45.6 & 139.3 &  1.49 & 0.10 & 4.8 \tblntmrk{e}\\
Abell~2219 & 0.226 & \ref{S99} & 16:40:20.7 & +46:42:39.8 &  O & 0.165 &\nd & \nd &  1 &16:40:22.0 &+46:42:47.0 & 15.4 &  14.87 & 0.17 & 239.1 \\
 & & & & &  O & 0.165 &  &  &  2 &16:40:23.6 &+46:42:15.3 & 38.9 &  0.97 & 0.17 & 7.9 \\
RXJ~1716+67 & 0.813 & \ref{H97} & 17:16:49.2 & +67:08:23.5 &  B & 0.109 & 30 & -34 &  1 &17:16:38.4 &+67:08:20.5 & 98.2 &  6.60 & 0.12 & 149.6 \\
 & & & & &  B & 0.109 &  &  &  2 &17:16:35.6 &+67:08:39.1 & 120.1 &  5.24 & 0.12 & 95.6 \\
 & & & & &  B & 0.109 &  &  &  3 &17:16:52.1 &+67:08:53.9 & 65.7 &  0.75 & 0.11 & \tblntmrk{g} \\
RX~J1720.1+2637 & 0.164 & \ref{E98} & 17:20:08.9 & +26:38:06.0 &  B & 0.167 &\nd & \nd &  1 &17:20:09.9 &+26:37:32.1 & 36.6 &  2.99 & 0.17 & 87.7 \\
 & & & & &  B & 0.167 &  &  &  2 &17:20:01.1 &+26:36:34.7 & 138.7 &  1.68 & 0.23 & 4.1 \\
Abell~2259 & 0.164 & \ref{S99} & 17:20:09.7 & +27:40:08.4 &  B & 0.131 & -10 &  1 &   & \nd & \nd & \nd & \nd & \nd & \nd \\
Abell~2261 & 0.224 & \ref{S99} & 17:22:27.1 & +32:07:58.6 &  B & 0.154 & -3 &  1 &  1 &17:22:17.0 &+32:09:12.6 & 144.8 &  9.32 & 0.22 & 23.0 \\
 & & & & &  O & 0.062 &  &  &  1 &17:22:17.1 &+32:09:14.1 & 144.8 &  10.48 & 0.16 & 23.0 \\
Abell~2294 & 0.178 & \ref{S99} & 17:23:55.3 & +85:53:24.0 &  B & 0.345 &\nd & \nd &   & \nd & \nd & \nd & \nd & \nd & \nd \\
MS~2053.7$-$0449 & 0.583 & \ref{G94} & 20:56:21.8 & -04:37:51.6 &  O & 0.031 & -13 & -4 &   & \nd & \nd & \nd & \nd & \nd & \nd \\
Abell~2345 & 0.177 & \ref{S99} & 21:27:13.6 & -12:09:45.4 &  O & 0.194 &\nd & \nd &  1 &21:27:09.7 &-12:10:00.9 & 59.9 &  5.07 & 0.22 & 57.6\tblntmrk{e} \\
 & & & & &  O & 0.194 &  &  &  2 &21:27:12.0 &-12:09:49.5 & 24.8 &  3.84 & 0.20 & 38.2\tblntmrk{e} \\
MACS~J2129.4$-$0741 & 0.570 & \ref{L03} & 21:29:26.0 & -07:41:28.0 &  O & 0.069 & -16 & -4 &  1 &21:29:30.2 &-07:42:30.2 & 98.0 &  1.18 & 0.10 & 0.6\tblntmrk{h} \\
RX~J2129.6+0005 & 0.235 & \ref{E98} & 21:29:37.9 & +00:05:38.4 &  B & 0.095 & 30 & -26 &  1 &21:29:39.9 &+00:05:21.8 & 9.4 &  2.33 & 0.10 & 25.4 \\
 & & & & &  B & 0.095 &  &  &  2 &21:29:55.1 &+00:08:01.7 & 284.2 &  3.68 & 0.48 & 34.3 \\
Abell~2409 & 0.148 & \ref{S99} & 22:00:54.5 & +20:57:32.4 &  B & 0.106 & -18 & 34 &  1 &22:01:11.3 &+20:54:55.6 & 316.4 &  3.40 & 0.46 & 3.9 \\
 & & & & &  B & 0.106 &  &  &  2 &22:01:18.0 &+20:57:51.9 & 346.9 &  4.47 & 0.85 & \tblntmrk{i} \\
MACS~J2214.9$-$1359 & 0.483 & \ref{L06} & 22:14:57.5 & -14:00:15.0 &  B & 0.159 & 14 &  4 &  1 &22:14:39.4 &-14:00:58.2 & 281.6 &  107.1 & 0.58 & 58.0 \\
 & & & & &  B & 0.159 &  &  &  2 &22:14:59.2 &-13:56:45.8 & 205.4 &  3.51 & 0.35 & 24.2 \\
 & & & & &  O & 0.128 &  &  &   & \nd & \nd & \nd &  \nd & \nd & \nd \\
MACS~J2228.5+2036 & 0.418 & \ref{2228} & 22:28:34.4 & +20:36:37.0 &  B & 0.110 & -20 & 37 &  1 &22:28:32.6 &+20:35:15.7 & 118.6 &  0.84 & 0.12 & 7.2 \\
 & & & & &  B & 0.110 &  &  &  2 &22:28:28.5 &+20:31:33.3 & 346.6 &  5.21 & 0.72 & 50.4 \\
MACS~J2243.3$-$0935 & 0.444 & \ref{2243} & 22:43:21.0 & -09:35:25.0 &  B & 0.618 & -35 & 12 &   & \nd & \nd & \nd & \nd & \nd & \nd \\
 & & & & &  O & 0.121 &  &  &  1 &22:43:17.8 &-09:35:08.4 & 12.5 &  0.93 & 0.13 & \tblntmrk{j} \\
Abell~2507 & 0.196 & \ref{S99} & 22:56:51.6 & +05:30:12.2 &  O & 0.072
&\nd & \nd &  1 &22:56:44.0 &+05:31:15.5 & 130.5 &  8.15 & 0.15 & 14.4
\\
\enddata
\tablenotetext{a}{No detection found in the literature. Using 3 times the NVSS noise level, we set an upper limit on the 1.4 GHz flux of 1.35 mJy and an upper limit on the 1.4 to 28.5 GHz spectral index of 0.0.}
\tablenotetext{b}{No detection found in the literature. Using 3 times the NVSS noise level, we set an upper limit on the 1.4 GHz flux of 1.35 mJy and an upper limit on the 1.4 to 28.5 GHz spectral index of 0.0.}
\tablenotetext{c}{FIRST catalog.}
\tablenotetext{d}{No detection found in the literature. Using 3 times the NVSS noise level, we set an upper limit on the 1.4 GHz flux of 1.35 mJy and an upper limit on the 1.4 to 28.5 GHz spectral index of 0.0.}
\tablenotetext{e}{VLA map. Fluxes are obtained by CLEANing the images in AIPS.}
\tablenotetext{f}{Using the 3 times the VLA map noise level, we set an upper limit on the 1.4 GHz flux of 0.45 mJy and an upper limit on the 1.4 to 28.5 GHz spectral index of -0.7.
The source is detected at 6~cm by \citet{partridge1986} with a flux of 2.59 mJy, which yields a spectral index of -0.4 from 5 to 28.5 GHz.}
\tablenotetext{g}{No detection found in the literature. Using the 3 times the VLA map noise level, we set an upper limit on the 1.4 GHz flux of 1.32 mJy and an upper limit on the 1.4 to 28.5 GHz spectral index of +0.2.}
\tablenotetext{h}{FIRST map.}
\tablenotetext{i}{No detection. Using the 3 times the NVSS noise level, we set an upper limit on the 1.4 GHz flux of 1.35 mJy and an upper limit on the 1.4 to 28.5 GHz spectral index of -0.4.}
\tablenotetext{j}{No detection. Using the 3 times the FIRST noise level, we set an upper limit on the 1.4 GHz flux of 0.45 mJy and an upper limit on the 1.4 to 28.5 GHz spectral index of -0.4.}
\tablecomments{
References:
\begin{enumerate}
\item \citet{stocke1991} \label{S91}
\item \citet{hughes1998} \label{H98}
\item From Chandra X-ray spectrum (observations 3521+5010, 45 ksec exposure) we obtain $z=0.586 \pm 0.01$. \label{0025}
\item \citet{dressler1999} \label{DRESS}
\item \citet{struble1999} \label{S99}
\item \citet{romer2000} \label{R00}
\item \citet{gladders2003} \label{G03}
\item \citet{gioia1994} \label{G94}
\item From Chandra X-ray spectrum (observations 3257+3582+6108, 70 ksec exposure) we obtain $z=0.467 \pm 0.005$. \label{0329}
\item \citet{ebeling1998} \label{E98}
\item \citet{laroque2003} \label{L03}
\item \citet{edge2003} \label{E03}
\item \citet{crawford1999} \label{C99}
\item \citet{dahle2002} \label{D02}
\item From Chandra X-ray spectrum (observation 509, 10 ksec exposure) we obtain $z=0.452 \pm 0.005$. \label{0913}
\item \citet{crawford1995} \label{C95}
\item \citet{allen1992} \label{A92}
\item \citet{luppino1995} \label{L95}
\item From Chandra X-ray spectrum (observations 3252+5009, 35 ksec exposure) we obtain $z=0.48^{+0.03}_{-0.06}$. \label{1108}
\item From Chandra X-ray spectrum (observations 3253+5008+5350, 35 ksec exposure) we obtain $z=0.51^{+0.04}_{-0.05}$. \label{1115}
\item \citet{bohringer2000} \label{B00}
\item \citet{donahue1999} \label{D99}
\item \citet{ebeling2001a} \label{E01}
\item From Chandra X-ray spectrum (observations 3258+6110, 85 ksec exposure) we obtain $z=0.519 \pm 0.007$. \label{1311}
\item \citet{gladders2005} \label{G05}
\item \citet{schindler1995} \label{S95}
\item \citet{patel2000} \label{P00}
\item \citet{henry1997} \label{H97}
\item \citet{laroque2006} \label{L06}
\item From Chandra X-ray spectrum (observation 3285, 20 ksec exposure) we obtain $z=0.42^{+0.01}_{-0.02}$. \label{2228}
\item From Chandra X-ray spectrum (observation 3260, 21 ksec exposure) we obtain $z=0.44 \pm 0.01$. \label{2243}
\end{enumerate}
}

\end{deluxetable}


\clearpage
\begin{deluxetable}{lllccccccccc}
\rotate
\tabletypesize{\scriptsize}
\tablecaption{ Radio Sources in Non-Cluster Fields. \label{tbl2}}
\tablewidth{0pt}
\tablehead{
\colhead{} &
\multicolumn{2}{c}{Pointing Center} &
\colhead{28.5 GHz} &
\multicolumn{5}{c}{28.5 GHz Source} &
\colhead{1.4GHz} \\
\colhead{} &
\multicolumn{2}{c}{\hrulefill} &
\colhead{RMS} &
\multicolumn{5}{c}{\hrulefill} &
\colhead{Flux}
\\
\colhead{Field} &
\colhead{$\alpha$(J2000)} &
\colhead{$\delta$(J2000)} &
\colhead{(mJy)} &
\colhead{$\alpha$(J2000)} &
\colhead{$\delta$(J2000)} &
\colhead{Radius ('')} &
\colhead{Flux (mJy)} &
\colhead{Unc. (mJy)} &
\colhead{(mJy)}
}
\startdata
BDF4 & 00:28:04.4 & +28:23:06.0 & 0.075 &  \nd & \nd & \nd & \nd & \nd & \nd \\
BDF4 & 00:28:04.4 & +28:23:06.0 & 0.115 &  \nd & \nd & \nd & \nd & \nd & \nd \\
HDF & 12:36:49.4 & +62:12:58.0 & 0.084 & 12:36:45.7 &+62:11:30.2 & 91.5 &  0.48 & 0.10 & \nd  \\
BDF6 & 18:21:00.0 & +59:15:00.0 & 0.074 &  \nd & \nd & \nd & \nd & \nd & \nd \\
BDF6 & 18:21:00.0 & +59:15:00.0 & 0.059 &  \nd & \nd & \nd & \nd & \nd & \nd \\
BDF7 & 06:58:45.0 & +55:17:00.0 & 0.083 &  \nd & \nd & \nd & \nd & \nd & \nd \\
BDF8 & 00:17:30.0 & +29:00:00.0 & 0.071 &  \nd & \nd & \nd & \nd & \nd & \nd \\
BDF9 & 12:50:15.0 & +56:52:30.0 & 0.084 &  \nd & \nd & \nd & \nd & \nd & \nd \\
BDF10 & 18:12:37.2 & +58:32:00.0 & 0.086 & 18:12:16.6 &+58:29:10.1 & 234.2 &  1.54 & 0.23 & \nd  \\
BDF11 & 06:58:00.0 & +54:24:00.0 & 0.090 &  \nd & \nd & \nd & \nd & \nd & \nd \\
BDF12 & 06:57:38.0 & +55:32:00.0 & 0.104 &  \nd & \nd & \nd & \nd & \nd & \nd \\
BDF13 & 22:22:45.0 & +36:37:00.0 & 0.097 &  \nd & \nd & \nd & \nd & \nd & \nd \\
\tableline
BDF14 & 00:26:04.4 & +28:23:06.0 & 0.093 &  \nd & \nd & \nd & \nd & \nd & \nd \\
BDF15 & 06:56:45.0 & +55:17:00.0 & 0.076 & 06:56:44.0 &+55:11:38.4 & 321.7 &  7.07 & 0.55 & 4.50  \\
BDF16 & 12:34:49.4 & +62:12:58.0 & 0.092 &  \nd & \nd & \nd & \nd & \nd & \nd \\
BDF17 & 18:19:00.0 & +59:15:00.0 & 0.090 &  \nd & \nd & \nd & \nd & \nd & \nd \\
BDF18 & 00:15:30.0 & +29:00:00.0 & 0.087 &  \nd & \nd & \nd & \nd & \nd & \nd \\
BDF19 & 06:55:38.0 & +55:32:00.0 & 0.084 &  \nd & \nd & \nd & \nd & \nd & \nd \\
BDF20 & 12:48:15.0 & +56:52:30.0 & 0.088 & 12:48:45.2 &+56:52:46.0 & 247.7 &  1.36 & 0.27 & 4.30  \\
BDF21 & 18:10:37.2 & +58:32:00.0 & 0.083 &  \nd & \nd & \nd & \nd & \nd & \nd \\
\enddata
\end{deluxetable}


\clearpage

\begin{deluxetable}{llcccc}
\rotate
\tabletypesize{\scriptsize}
\tablecaption{Spectral indices. \label{tbl3}}
\tablewidth{0pt}
\tablehead{\colhead{Data Set}  &  \colhead{Frequency Range} &  \colhead{No.\ Sources} &  \colhead{Mean Index $\pm$ RMS} &   \colhead{Median Index [25\%, 75\%]} &   \colhead{Flux Limits} }
\startdata
Overall cluster                       & 1.4 to 28.5 GHz    & 88  & 0.66 $\pm$ 0.36              & 0.72 [   0.51, 0.92]   & $\sim$ 0.6 - 10.0 mJy at 28.5 GHz (see text)\\
Inner cluster ($r \leq 0.5$ arcmin)   & 1.4 to 28.5 GHz    & 21  & 0.75 $\pm$ 0.24              & 0.76 [   0.56, 0.94]   & $\sim$ 0.6 - 10.0 mJy at 28.5 GHz (see text)\\
Outer cluster ($r \geq 0.5$ arcmin)   & 1.4 to 28.5 GHz    & 67  & 0.63 $\pm$ 0.38              & 0.71 [   0.42, 0.88]   & $\sim$ 0.7 - 8.0 mJy at 28.5 GHz (see text)\\
\tableline
Mason et al. 2003 (CBI)               & 1.4 to 31 GHz      & 56  & 0.45 $\pm$ 0.37              & \nd                    & 21 mJy at 31 GHz\\
Bolton et al. 2004 (9C follow-up)     & 1.4 to 4.8 GHz     & 124 & \nd                          & 0.44 [   0.05, 0.76]   & 25 mJy at 15.2 GHz\\
Bolton et al. 2004 (9C follow-up)     & 4.8 to 15.2 GHz    & 124 & \nd                          & 0.39 [   0.06, 0.95]   & 25 mJy at 15.2 GHz\\
Bolton et al. 2004 (9C follow-up)     & 15.2 to 43 GHz     & 124 & \nd                          & 0.87 [   0.42, 1.20]   & 25 mJy at 15.2 GHz\\
Bolton et al. 2004 (9C follow-up)     & 1.4 to 4.8 GHz     & 70  & \nd                          & 0.24 [$-$0.12, 0.64]   & 60 mJy at 15.2 GHz\\
Bolton et al. 2004 (9C follow-up)     & 4.8 to 15.2 GHz    & 70  & \nd                          & 0.27 [   0.02, 0.70]   & 60 mJy at 15.2 GHz\\
Bolton et al. 2004 (9C follow-up)     & 15.2 to 43 GHz     & 70  & \nd                          & 0.67 [   0.38, 1.03]   & 60 mJy at 15.2 GHz\\
\enddata
\end{deluxetable}


\clearpage

\begin{deluxetable}{cccccc}
\rotate
\tabletypesize{\scriptsize}
\tablecaption{Number Counts for $\geq 5 \sigma$ Sources at $28.5$ GHz \label{tbl4}}
\tablewidth{0pt}
\tablehead{\colhead{log10(S) (mJy)}  &  \colhead{radius range (arcmin)\tablenotemark{a}} &  \colhead{raw counts} &  \colhead{area (arcmin$^{2}$)}  & \colhead{$log_{10}(dN/dS)$ (arcmin$^{-2}$ mJy$^{-1}$)} & \colhead{field type}}
\startdata
$-$0.15 to 0.00 & $\geq 0.5$    & 10  &  783 & $-1.35^{+0.15}_{-0.16}$ 	& cluster\\
0.00 : 0.15    & $\geq 0.5$    & 6   & 1541 & $-2.03^{+0.20}_{-0.22}$ 	& cluster\\
0.15 : 0.30    & $\geq 0.5$    & 10  & 2420 & $-2.15^{+0.15}_{-0.15}$ 	& cluster\\
0.30 : 0.45    & $\geq 0.5$    & 6   & 3306 & $-2.66^{+0.20}_{-0.22}$ 	& cluster\\
0.45 : 0.60    & $\geq 0.5$    & 11  & 4217 & $-2.65^{+0.15}_{-0.15}$ 	& cluster\\
0.60 : 0.75    & $\geq 0.5$    & 8   & 5104 & $-3.02^{+0.17}_{-0.18}$ 	& cluster\\
0.75 : 0.90    & $\geq 0.5$    & 7   & 5951 & $-3.29^{+0.19}_{-0.20}$ 	& cluster\\
\tableline
$-$0.20 to 0.10 & $\leq 0.5$    & 4   & 42.0 & $-0.82^{+0.25}_{-0.28}$	& cluster\\
0.10 : 0.40    & $\leq 0.5$    & 6   & 63.4 & $-1.12^{+0.20}_{-0.22}$ 	& cluster\\
0.40 : 0.70    & $\leq 0.5$    & 3   & 69.0 & $-1.76^{+0.30}_{-0.34}$ 	& cluster\\
0.70 : 1.00    & $\leq 0.5$    & 2   & 69.5 & $-2.24^{+0.37}_{-0.45}$ 	& cluster\\
\tableline
0.13 : 0.50    & $\geq 0.0$    &  1  &  438 & $-2.90^{+0.52}_{-0.76}$	& non-cluster\\
0.50 : 0.87    & $\geq 0.0$    &  1  &  724 & $-3.49^{+0.52}_{-0.76}$	& non-cluster\\
\enddata
\tablenotetext{a}{Max radius varies for individual fields since set by noise. See text.}
\end{deluxetable}


\clearpage

\begin{deluxetable}{lcccc}
\rotate
\tabletypesize{\scriptsize}
\tablecaption{Power-law Fits to $dN/dS$ \label{tbl5}}
\tablewidth{0pt}
\tablehead{\colhead{Field type}  &  \colhead{Best indiv. fit index} &  \colhead{Best indiv. fit normalization} &  \colhead{Best joint fit index}&  \colhead{Best joint fit normalization}}
\startdata
Inner cluster ($r \leq 0.5$ arcmin) & $-$1.67 $\pm$ 0.37 & $132^{+108}_{-41}    \times 10^{-3}$ mJy$^{-1}$ arcmin$^{-2}$ & $-$1.98 $\pm 0.20$ &  $174^{+60}_{-45}    \times 10^{-3}$ mJy$^{-1}$ arcmin$^{-2}$\\
Outer cluster ($r \geq 0.5$ arcmin) & $-$2.02 $\pm$ 0.22 & $20.9^{+5.4}_{-4.3} \times 10^{-3}$ mJy$^{-1}$ arcmin$^{-2}$ & $-$1.98 $\pm 0.20$ & $19.5^{+4.0}_{-5.0}  \times 10^{-3}$ mJy$^{-1}$ arcmin$^{-2}$\\
Non-cluster                         & \nodata & \nodata                                                      & $-$1.98 $\pm 0.20$ & $5.90^{+6.70}_{-3.15} \times 10^{-3}$ mJy$^{-1}$ arcmin$^{-2}$\\
\enddata
\end{deluxetable}


\end{document}